\documentstyle[12pt,a4,epsf]{article}
\newcommand{\lsim}{\buildrel < \over {_\sim}}
\newcommand{\ie}{{\it i.e.}}
\newcommand{\eg}{{\it e.g.}}
\newcommand{\cf}{{\it cf.}}

\newcommand{\qu}{{\rm q}}
\newcommand{\qb}{${\rm\bar q}$}
\newcommand{\qbm}{{\rm\bar q}}

\newcommand{\pvec}{\vec p}
\newcommand{\kvec}{\vec k}
\newcommand{\rvec}{\vec r}
\newcommand{\Rvec}{\vec R}
\newcommand{\lqcd}{\Lambda_{QCD}}
\newcommand{\ieps}{i\varepsilon}
\newcommand{\disc}{{\rm Disc}}
\newcommand{\M}{{\cal M}}
\newcommand{\pl}{{||}}

\newcommand{\order}[1]{${\cal O}\left(#1 \right)$}
\newcommand{\morder}[1]{{\cal O}\left(#1 \right)}
\newcommand{\eq}[1]{(\ref{#1})}

\newcommand{\beq}{\begin{equation}}
\newcommand{\eeq}{\end{equation}}
\newcommand{\beqa}{\begin{eqnarray}}
\newcommand{\eeqa}{\end{eqnarray}}

\newcommand{\ket}[1]{\vert{#1}\rangle}

\newcommand{\PL}[3]{Phys.~Lett.~{\bf {#1}},~{#2}~({#3})}
\newcommand{\NP}[3]{Nucl.~Phys.~{\bf {#1}},~{#2}~({#3})}
\newcommand{\PRD}[3]{Phys.~Rev.~{\bf D{#1}},~{#2}~({#3})}
\newcommand{\PRL}[3]{Phys.~Rev.~Lett.~{\bf {#1}},~{#2}~({#3})}

\newcommand{\PRe}[3]{Phys.~Rep.~{\bf {#1}},~{#2}~({#3})}

\renewcommand{\thefootnote}{\fnsymbol{footnote}}

\begin{document}

\begin{titlepage}
\begin{flushright}
           \today\\
           LAPTH-842/01\\
           NORDITA-2001-7 HE\\
           SLAC-PUB-8818\\
           hep-ph/0104291\\
\end{flushright}

\vskip 1.5cm

   \centerline{{\large STRUCTURE FUNCTIONS ARE NOT PARTON
PROBABILITIES\footnote{This work has been supported in part by the US
Department
of Energy under contract DE-AC03- 76SF00515 (SJB) and by the EU
Commission under
contracts HPRN-CT-2000-00130 (PH and FS) and HPMT-2000-00010 (NM).}}}

\vskip 1.cm

\centerline{ Stanley J. Brodsky$^1$, Paul Hoyer$^{2,}$\footnote{On leave of
absence from the Department of Physics, University of Helsinki, Finland.},}
\vspace{.2cm}
\centerline{ Nils Marchal$^{2,3}$, St\'ephane Peign\'e$^3$ and Francesco
Sannino$^2$}

\vskip .5cm

{\small\sl
\centerline{ $^1$Stanford Linear Accelerator Center, Stanford CA 94309,
USA}
\vspace{.1cm}
\centerline{ $^2$Nordita, Blegdamsvej 17, DK--2100 Copenhagen, Denmark}
\vspace{.1cm}
\centerline{ $^3$LAPTH\footnote{CNRS, UMR 5108, associated to the
      University of Savoie.}, BP 110, F-74941 Annecy-le-Vieux Cedex, France}}

\vskip 1cm

\begin{abstract}
The common view that structure functions measured in deep
inelastic lepton scattering are determined by the {\it
probability} of finding quarks and gluons in the target is not
correct in gauge theory. We show that gluon exchange between the
fast, outgoing partons and target spectators, which is usually
assumed to be an irrelevant gauge artifact, affects the leading
twist structure functions in a profound way. This observation
removes the apparent contradiction between the projectile
(eikonal) and target (parton model) views of diffractive and small
$x_B$ phenomena. The diffractive scattering of the fast outgoing
quarks on spectators in the target causes shadowing in the DIS
cross section. Thus the depletion of the nuclear structure
functions is not intrinsic to the wave function of the nucleus,
but is a coherent effect arising from the destructive interference
of diffractive channels induced by final state interactions. This
is consistent with the Glauber-Gribov interpretation of shadowing
as a rescattering effect.

\end{abstract}

\end{titlepage}

\newpage
\renewcommand{\thefootnote}{\arabic{footnote}}
\setcounter{footnote}{0}
\setcounter{page}{1}

{\bf 1. Introduction}

Deep inelastic lepton scattering, $\ell N \to \ell' + X$ (DIS) is
central for our understanding of hadron structure. Ever since the
earliest days of the parton model, it has been assumed that the
leading-twist structure functions $F_i(x,Q^2)$ measured in deep
inelastic lepton scattering are determined by the {\it
probability} to find quarks and gluons in the target \cite{dy}.
This probability is given by the target wave function at the
light-cone (LC) time when the current interacts (in the $q^+ \le
0$ frame). For example, the quark probability is distribution is
\beq {\cal P}_{\qu/N}(x_B,Q^2)= \sum_n
\int^{k_{i\perp}^2<Q^2}\left[ \prod_i\, dx_i\, d^2k_{\perp
i}\right] |\psi_n(x_i,k_{\perp i})|^2 \sum_{j=q} \delta(x_B-x_j)
\label{strfn} \eeq where the $\psi_n$ are LC wave functions of the
target (see Eq. \eq{fock} below). The identification of structure
functions with the square of light-front wave functions is usually
made in the ghost-free LC gauge $n\cdot A = A^+=0$, the argument
being that the path-ordered exponential in the operator product
appearing in parton distributions (see Eq. \eq{melm} below)
reduces to unity. Thus the DIS cross section appears to be fully
determined by the probability distribution of partons in the
target.

However, we shall show that this parton model interpretation of
the structure functions, which was established for a theory with
Yukawa couplings~\cite{dy}, is not correct in gauge theory. The
critical issue is whether the scattering taking place after the
virtual photon interacts can affect the leading twist cross
section. It is well-known that in Feynman and other covariant
gauges one has to include corrections to the ``handbag" diagram
due to final state interactions of the struck quark  with the
gauge field of the target. Light-cone gauge is singular -- in
particular, the gluon propagator $ d_{LC}^{\mu\nu}(k) =
\frac{i}{k^2+\ieps}\left[-g^{\mu\nu}+\frac{n^\mu k^\nu+ k^\mu
n^\nu}{n\cdot k}\right]  $ has a pole at $k^+ = 0$ which requires
an analytic prescription.  In final-state scattering involving
on-shell intermediate states, the exchanged momentum $k^+$ is of
\order{1/\nu} in the target rest frame, which enhances the second
term in the propagator.  This enhancement allows rescattering to
contribute at leading twist even in light-cone gauge.

We find that gluon exchange between the outgoing quarks and
target spectators, which is usually assumed to be suppressed in the
Bjorken limit, affects the leading twist structure functions in a profound
way. Final state diffractive scattering gives rise to
interference effects in the DIS cross section. Thus nuclear shadowing is
not caused by the wave function of the nucleus, but is induced by final
state interactions.

Thus the depletion of the nuclear structure
functions is not intrinsic to the wave function
of the nucleus, but in
fact is a coherent effect reflecting the destructive interference of
diffractive channels induced by the final state interactions.

The
distinction between structure functions and parton probabilities is
already implied by the Glauber-Gribov picture of nuclear
shadowing~\cite{Gribov:1969jf,Brodsky:1969iz,Brodsky:1990qz,Piller:2000wx}.
In this framework shadowing arises from interference between complex
rescattering amplitudes involving on-shell intermediate states.  In
contrast, the
wave function of a stable target is strictly real since it does not have on
energy-shell configurations.  A probabilistic interpretation of the DIS cross
section is thus precluded.

Our paper thus explains the origins of nuclear
shadowing  and leading-twist diffraction, giving a new, first principle,
perspective on these problems. Our formalism of
final-state interactions has recently been used to analyze
single-spin asymmetries
in deep inelastic processes and to show that such asymmetries survive in the
Bjorken limit, contrary to conventional arguments which claim that
final state interactions are always power-law suppressed in the
large scale of hard QCD processes~\cite{Brodsky:2002cx}.

\vspace{.5cm}

{\bf 2. The Foundations of the QCD-Improved Parton Model}

Soon after the observation of
Bjorken scaling (and before the advent of QCD) it was suggested \cite{dy} that
the DIS cross section is fully determined by the target wave function.
Specifically, consider the Fock expansion of the nucleon state $\ket{N}$
in terms of its quark and gluon constituents at equal Light-Cone (LC) time
$\tau=t+z/c=y^+ = y^0+y^3$,
\beqa
\ket{N}&=& \int \left[\prod_i\, \frac{dx_i\, d^2\kvec_{\perp
i}}{16\pi^3}\right]
\Big[\psi_{uud}(x_i,\kvec_{\perp i},\lambda_i) \ket{uud} \nonumber \\
&+& \psi_{uudg}(\ldots) \ket{uudg}+ \ldots + \psi_{\cdots}(\ldots)
\ket{uudq\bar q}+ \ldots \Big] \label{fock}
\eeqa
Each Fock state $\ket{uud\ldots}$ is weighted by an amplitude $\psi$
which depends on the LC momentum fractions $x_i=k_i^+/p^+$
($\sum_i x_i = 1$),
the relative transverse momenta $\kvec_{\perp i}$ ($\sum_i \kvec_{\perp i}
= 0$),
and the helicities $\lambda_i$ of its constituents\footnote{See Ref.
\cite{spd} for the normalization conventions.}. The DIS cross section
thus appeared to measure
the single parton probabilities ${\cal P}_{j/N}(x_B,Q^2)$ as defined in
\eq{strfn}, which
express the probability  for finding
(at resolution $1/Q$) a parton $j$ carrying the momentum fraction
$x_B=Q^2/2p\cdot q$ of the nucleon. Here $q$ is the virtual photon momentum
$(q^2=-Q^2)$ and $p$ the target nucleon momentum.

Later analyses \cite{css} of perturbative QCD (PQCD) have
established the QCD factorization theorem to all orders in the coupling. The
DIS cross section can be expressed for each parton type as a convolution
of a perturbatively calculable hard subprocess cross section and a
target parton
distribution. The parton distributions are given by operator matrix elements
of the target. For the (spin-averaged) quark distribution in the nucleon
$N$ of momentum $p$,
\beqa
f_{\qu/N}(x_B,Q^2)&=& \frac{1}{8\pi} \int dy^- \exp(-ix_B p^+ y^-)
\label{melm}\\
&\times&\langle N(p)| \qbm(y^-) \gamma^+\, {\rm P}\exp\left[ig\int_0^{y^-}
dw^- A^+(w^-) \right] \qu(0)|N(p)\rangle \nonumber
\eeqa
where all fields are evaluated at equal LC time $y^+ = 0$ and vanishing
transverse separation $y_\perp=0$. The light-like distance between the
absorption and emission vertices of the virtual photon in the forward
amplitude is measured by $y^-$. The path-ordering ${\rm P}$ orders the
gauge fields according to their position on the light-cone and ensures the
gauge invariance of the matrix element.

The identification of the quark distribution \eq{melm} as a probability
distribution \eq{strfn} is made in LC gauge $n\cdot A = A^+=0$, where
the path-ordered exponential in \eq{melm} reduces to unity, and one finds
$f_{\qu/N} \to {\cal P}_{\qu/N}$. A recent derivation in the more general
case of non-forward matrix elements (Skewed Parton Distributions) may be
found in Ref. \cite{spd}. Thus the DIS cross section appears to be fully
determined by the probability distribution of partons in the target.
However, as we shall show
the expression for $f_{\qu/N}$ cannot be given by \eq{melm} in LC gauge.

In a general gauge the matrix element \eq{melm} depends on final state
interactions (FSI) of the struck quark with the gauge field of the target
via the $A^+$-dependence of the path-ordered exponential. Based on the above
argument in LC gauge, it is generally believed that the exponential is a
gauge artifact  and thus that the presence of FSI
does not influence the cross section.
But this assumes that $f_{\qu/N}$ is given by \eq{melm} in all gauges,
{\it including} LC gauge.
Here we find that final state rescattering
in fact {\it does} change the DIS cross section
in {\it all} gauges.
Our analysis is
consistent with the QCD factorization theorem and with the form \eq{melm} of
the parton distributions in all gauges except LC gauge.

The influence of FSI we find at leading-twist is specific to gauge
theories. The impossibility to interpret parton distributions as
probabilities could thus not be inferred before the advent of QCD.
Instead, the equivalence between DIS structure functions and the
target wave function was assumed, though it was only shown in a theory
with Yukawa coupling \cite{dy}.

The expression \eq{melm} for $f_{\qu/N}$ is
valid for covariant gauges in the Bjorken limit,
which selects the $A^+$ field of the target. We shall show that setting then
$A^+=0$ in \eq{melm} leads to an incorrect expression for $f_{\qu/N}$.
  From a mathematical point of view this means that the high
energy Bjorken limit does not commute with the $A^+ \to 0$ limit of
LC gauge.
In fact (see section 7) the high energy and LC gauge limits do
not commute even for ordinary elastic electron scattering.

In section 3 we recall why in Feynman gauge
final state interactions among the
spectator partons of the target system do not affect the DIS cross
section at leading twist. We then show that this general argument
does not apply to rescattering of the struck quark.

In section 4 we discuss the Glauber-Gribov picture and show why it
implies that the final state interactions,
resummed in covariant gauges by the path ordered exponential of \eq{melm},
affect the cross section. We then study a simple perturbative model of
rescattering effects in section 5, for which explicit expressions of the
amplitudes can be obtained at small $x_B$. Using this example we
demonstrate in section 6 that rescattering of the struck quark on the
target can cause a leading twist shadowing effect.

The analysis of sections 3 to 6 is carried out in Feynman gauge.
In section 7 we show why rescattering effects can persist even in
$A^+=0$ gauge, in contradiction with the form \eq{melm} of the
matrix element. As is well-known, this gauge is singular -- in
particular, the gluon propagator
\beq d_{LC}^{\mu\nu}(k) =
\frac{i}{k^2+\ieps}\left[-g^{\mu\nu}+\frac{n^\mu k^\nu+ k^\mu
n^\nu}{n\cdot k}\right]  \label{lcprop} \eeq
has a pole at $n\cdot k = k^+ =0$
which requires an analytic prescription. In
final-state elastic scattering of the struck quark the exchanged momentum
$k^+$ is of
\order{1/\nu} in the target rest frame, which enhances the second
term in the propagator \eq{lcprop}. This enhancement allows
rescattering to contribute at leading twist in LC gauge.

We reevaluate our model amplitudes using LC gauge in the Appendix. Although the
expressions for the individual diagrams depend on the prescription used at
$n\cdot k=0$, the prescription dependence vanishes when all diagrams are added.
The scattering amplitudes which we calculate up to two-loops in LC gauge thus
agree with the result in Feynman gauge.

For the issues of this paper, the spin and color of the quarks are
not relevant. We therefore conduct our discussion in the simpler framework of
abelian gauge theory with scalar quarks.

\vspace{.5cm}

{\bf 3. Effects of final state interactions in deep inelastic scattering}

The DIS cross section is given by the discontinuity of the forward amplitude,
\beq
\sigma(\gamma^* T \to X) = \frac{1}{4M\nu} \disc \M(\gamma^* T \to \gamma^* T)
\label{optical}
\eeq
where $M$ is the target mass and $\nu$ the photon energy in the rest system
of the target. We take the Bjorken limit $\nu, Q^2 =-q^2 \to \infty$ with
$x_B=Q^2/2M\nu$ fixed. In the LC notation $k=(k^+,k^-,\kvec_\perp)$, where
$k^{\pm}=k^0\pm k^3$, the photon and target momenta are (at leading order)
\beqa
q &=& (-Mx_B,2\nu,0_\perp) \nonumber\\
p &=& (M,M,0_\perp)  \label{itrframe}
\eeqa

In the following we define a final state interaction (FSI) as any interaction
which occurs after the virtual photon has been absorbed. Here `after' refers to
LC time, $y^+=y^0+y^3$,  in the frame \eq{itrframe}. In deep inelastic
scattering
initial state interactions (ISI) occur only within the target bound state and
determine the target wave function \eq{fock}. We shall show that soft
rescattering of the struck quark in the target also affects the DIS cross
section.

We can distinguish FSI from ISI using LC time-ordered perturbation
theory (LCPT) \cite{lb}.  Fig. 1 illustrates two LCPT diagrams
which contribute to the forward $\gamma^* T \to \gamma^* T$
amplitude, where the target $T$ is taken to be a single quark. We
use these diagrams in a generic sense here, while in sections 5
and 6 we consider them in the framework of a specific perturbative
model of the DIS process.

\begin{figure}[hbt]
\begin{center}
\leavevmode
{\epsfxsize=14truecm \epsfbox{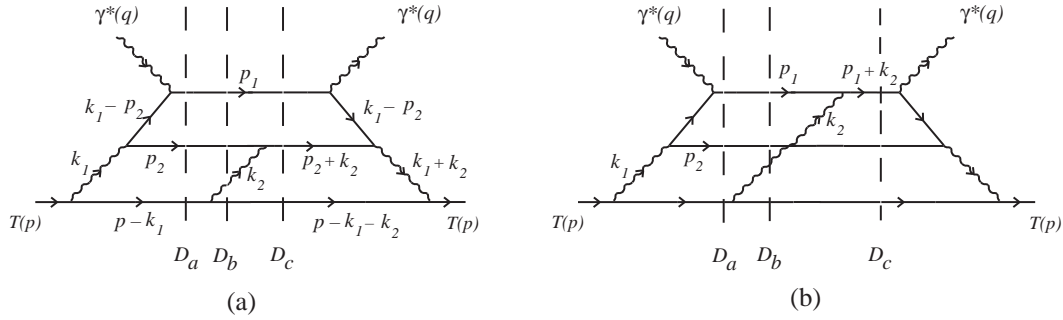}}
\end{center}
\caption[*]{Two types of final state interactions. (a) Scattering of the
antiquark ($p_2$ line), which in the aligned jet kinematics is part of the
target dynamics. (b) Scattering of the current quark ($p_1$ line). For each LC
time-ordered diagram, the potentially on-shell intermediate states
corresponding to the denominators $D_a, D_b, D_c$ are denoted by dashed
lines.}
\label{fig1dis}
\end{figure}

We recall that in LCPT the `$-$' momentum component is not an independent
variable, but is given by the on-shell condition, $k^-=(k_{\perp}^2+m^2)/k^+$.
Each propagating line has a factor $1/k^+$, and there is a denominator
factor
\beq
D_{int}=\sum_{inc}k^- - \sum_{int}k^- +\ieps  \label{dint}
\eeq
for each intermediate state, which measures the LC energy difference
between the
incoming and intermediate states. In Feynman gauge (which we use in this
section)
an imaginary part or discontinuity can arise only via the $\ieps$ prescription
in \eq{dint}, when LC energy is conserved and the intermediate state is
on-shell.

We consider the `aligned jet' (or parton model) configuration \cite{bks},
where the hard vertex is taken at zeroth order in the strong coupling:
$\gamma^* \qu \to \qu$. In the aligned jet kinematics the momentum $p_1$ of
the struck quark in Fig. 1 is the only one which grows in the Bjorken limit:
$p_1^- \simeq 2\nu$, with $\pvec_{1\perp}$ independent of $\nu$. All momenta in
Fig.~1 other than $q$ and $p_1$ remain finite in the Bjorken limit. The
condition that the momentum fraction of the struck quark equals $x_B$ follows
  from the conservation of `$+$' momentum, given that $p_1^+ =
\morder{1/\nu}$.

We recall (see, \eg, Eq. (A5) of Ref. \cite{bhm}) that the virtual photon
polarization vectors may be chosen as
\beqa
\varepsilon(\lambda=\pm 1) &=& -\frac{1}{\sqrt{2}}(0,0,1,\pm i)
\nonumber\\
\varepsilon(\lambda=0) &=& \frac{Q}{\nu}(1,-1,0,0)  \label{polvec}
\eeqa
Since we take all lines (except the gauge bosons) in Fig. 1 to be scalars, the
longitudinal photon coupling $\varepsilon(\lambda=0)\cdot (p_1+k_1-p_2)
\simeq Q$ dominates over the transverse ones in the Bjorken limit. The two
longitudinal photon couplings together contribute a factor $Q^2$ to the
forward amplitudes in Figs. 1a and 1b.

Both diagrams in Fig. 1 contain final state interactions between the
$\gamma^*$ vertices. Only the three intermediate states indicated by dashed
vertical lines can  kinematically be on-shell and thus contribute to the
discontinuity of the diagrams via the vanishing of the corresponding
denominator $D_a,\ D_b$ or
$D_c$. We wish to ascertain whether the sum of these discontinuities gives a
leading-twist contribution to the DIS cross section through the optical
theorem \eq{optical}.
We use Feynman gauge in the following discussion. As we shall
see in section 7 and Appendix C, the specific Feynman diagrams causing
FSI effects in DIS actually depend on the gauge.

The three denominators of Fig. 1a are
\beqa
D_a&=& q^- +p^- -p_1^- -p_2^- -(p-k_1)^- \nonumber\\
&=& 2\nu - \frac{p_{1\perp}^2+m^2}{p_1^+} + M
-\frac{p_{2\perp}^2+m^2}{p_2^+} - \frac{k_{1\perp}^2+M^2}{M-k_1^+}
\nonumber\\ && \label{denoms}\\
D_b &=& 2\nu - \frac{p_{1\perp}^2+m^2}{p_1^+} + M
-\frac{p_{2\perp}^2+m^2}{p_2^+} -\frac{k_{2\perp}^2}{k_2^+} -
\frac{(\kvec_{1\perp}+\kvec_{2\perp})^2+M^2}{M-k_1^+-k_2^+} \nonumber\\
&&\nonumber\\
D_c&=& 2\nu - \frac{p_{1\perp}^2+m^2}{p_1^+} + M
-\frac{(\pvec_{2\perp}+\kvec_{2\perp})^2+m^2}{p_2^+ +k_2^+}
-\frac{(\kvec_{1\perp}+\kvec_{2\perp})^2+M^2}{M-k_1^+ -k_2^+} \nonumber
\eeqa
and have the form
\beq
D_{a,b,c} = 2\nu - \frac{p_{1\perp}^2+m^2}{p_1^+} + f_{a,b,c}
\label{Dcancel}
\eeq
where $f_a,f_b,f_c$ are independent of $\nu$ in the aligned jet configuration.
If we consider these denominators as functions of $p_1^+$ then the three
conditions
$D_{a,b,c}=0$ give to leading order the {\em same} value of $p_1^+$,
\beq
p_1^+ = \frac{p_{1\perp}^2+m^2}{2\nu}\left[1+\morder{\frac{1}{\nu}}\right]
\label{poneplus}
\eeq

All denominators and other factors in the LCPT expression of
Fig.~1a except $D_a, D_b$ and $D_c$  are insensitive (at leading
order) to a relative change in $p_1^+$ of \order{1/\nu}. Thus, as
far as the discontinuity of Fig.~1a is concerned, we can regard
the other factors as constants in the $p_1^+$-integral containing
the denominator poles, \beq \mbox{Disc(Fig. 1a)} \propto Q^2 \,
\disc\int \frac{dp_1^+}{p_1^+}
\frac{1}{(D_a+\ieps)(D_b+\ieps)(D_c+\ieps)} \label{impart} \eeq
where the factor $Q^2$ stems from the photon couplings. All
remaining factors in the proportionality are independent of $\nu$.
Each of the three denominators in \eq{impart} gives a
$\nu$-independent contribution to the discontinuity in the Bjorken
limit. This means that each partial discontinuity contributes to
the DIS cross section of Eq. \eq{optical} at the leading twist
level, $\sigma(\gamma^* T \to X) \propto 1/Q^2$. However, as is
easily seen, the contributions from the three poles which to
leading order occur at the same value \eq{poneplus} of $p_1^+$
cancel at leading twist.

The above argument is generic and applies to arbitrarily complex
diagrams having
no interactions on the current quark line $p_1$. The remarkable fact that FSI
between target spectators do not affect the DIS cross section only
relies on the
Bjorken limit, which as $\nu\to \infty$ provides an `infinite energy reservoir'
which compensates any target excitations.

The situation is quite different for diagrams like Fig.~1b where the current
quark reinteracts. In (quasi-)elastic scattering of the current quark the
momentum transfer $k_2^+ \propto 1/\nu$. We may check explicitly that
this range
of momentum transfer indeed gives a leading-twist contribution to each partial
discontinuity. The denominators are now of the form
\beqa
D_a &\simeq& 2\nu- \frac{p_{1\perp}^2+m^2}{p_1^+} +g_a \nonumber\\
&& \nonumber\\
D_b &\simeq& 2\nu- \frac{p_{1\perp}^2+m^2}{p_1^+} -\frac{k_{2\perp}^2}{k_2^+}
+g_b  \label{denomsbis} \\
&& \nonumber\\
D_c &\simeq& 2\nu- \frac{(\pvec_{1\perp}+\kvec_{2\perp})^2+m^2}{p_1^+ +k_2^+}
+g_c \nonumber
\eeqa
where $g_{a,b,c}$ are again independent of $\nu$. For example, picking up the
$D_a=0$ pole in the $p_1^+$ integral we have
\beq
\mbox{Disc$_a$(Fig. 1b)} \propto Q^2 {p_1^+}{p_1^-}\, \int
\frac{dk_2^+}{k_2^+ (p_1^+ +k_2^+)} \frac{1}{D_b\,D_c} \label{impartb}
\eeq
where
$p_1^+$ is given by \eq{poneplus} and the factor $p_1^-$  originates from the
interaction in Feynman gauge.

Note that $D_b$ and $D_c$ are still of \order{\nu} at the value
\eq{poneplus} of $p_1^+$ for which $D_a=0$. The fact that the
contributions from $D_a=0,\ D_b=0$ and $D_c=0$ thus occur at
distinct values of $p_1^+$ means that they no longer cancel.
$\disc_a$ is independent of $\nu$ and hence contributes to the DIS
cross section at leading twist. We conclude that rescattering of
the current quark generally affects the cross section. In section
6 we shall demonstrate, in terms of an explicit perturbative
example, that this conclusion is indeed correct.

Since the LC energy differences $D_{b,c} \propto \nu$ at $D_a=0$,
the struck quark rescattering occurs on the light-cone, $y^+
\sim\morder{1/\nu}$. This rescattering is part of the dynamics
described by the path-ordered exponential in the matrix element
\eq{melm}, where all $A^+$ fields are evaluated at the same LC
time $y^+$. During its passage through the target the struck quark
has no time to emit or absorb gluons, it only `samples' the
Coulomb field of the target. The rescattering nevertheless changes
the transverse momentum of the quark and influences the cross
section. This is analogous to the LPM effect \cite{lpm}, which
suppresses the bremsstrahlung of a high energy electron in matter
due to Coulomb rescattering within the formation time of the
radiated photons.

\vspace{.5cm}

{\bf 4. The Glauber-Gribov Picture of Shadowing}

DIS data on nuclear targets $A$ has shown that nuclear structure functions
are suppressed for $x_B \lsim 0.05$: $F_2^A(x_B,Q^2) < A\,F_2^N(x_B,Q^2)$
\cite{pw}. This is generally interpreted as a leading twist `shadowing' effect,
arising from quantum mechanical interference \cite{gribov,pw}. The coherence
length of the virtual photon in the target rest frame \eq{itrframe} is long at
small $x_B$,
\beq
   \frac{2\nu}{Q^2} = \frac{1}{Mx_B} = \langle y^- \rangle   \label{ioffe}
\eeq as can also be seen from Eq. \eq{melm}. Rescattering from
different nucleons in the nucleus can thus interfere.

In the aligned jet kinematics the virtual photon fluctuates into a \qu\qb\ pair
with limited transverse momentum, and the (struck) quark takes nearly all the
longitudinal momentum of the photon. Using the notation of Fig. 1, where the
initial \qu\ and \qb\ momenta are denoted $p_1$ and $p_2-k_1$, respectively, we
have
\beqa \label{alkin}
p_1^- &\simeq& 2\nu \nonumber \\
p_2^+ - k_1^+ &\simeq& -Mx_B \\
\pvec_{1\perp} = -(\pvec_{2_\perp} - \kvec_{1_\perp}) &\sim& \lqcd \nonumber
\eeqa
The (covariant) virtualities $p_1^2$ and $(p_2-k_1)^2$ are limited. Hence
$(p_1+p_2-k_1)^2 \sim p_1^- (p_2^+ - k_1^+) \sim -Q^2$ as required by momentum
conservation. The virtual quark pair is put on-shell by a (total) momentum
transfer $k$ from the target, with
\beq \label{kexch}
k^+ = p_1^+ +p_2^+-q^+ \simeq p_2^+ + Mx_B
\eeq
The DIS cross section is dominated by minimal transfers $k^+$, which for
the final antiquark momentum gives
\beq \label{p2plus}
p_2^+ \sim Mx_B
\eeq

With this kinematics in mind the Glauber-Gribov picture of shadowing can be
summarized as follows. At small $x_B$ the antiquark momentum $p_2^-
\propto 1/x_B$ is large but the momentum transfer $k^+ \sim Mx_B$ is small.
The scattering will therefore have a diffractive component. In particular, the
quark pair may scatter elastically on a `front' nucleon $N_1$ in the nucleus
before suffering an inelastic collision at a `back' nucleon $N_2$, as
indicated on the lhs of Fig. 2. The small momentum transfer $k^+$ at $N_1$
required to put the quark pair on-shell can be absorbed by the nuclear wave
function. Hence this amplitude interferes with the amplitude for a single
scattering on $N_2$ shown on the rhs of Fig. 2. The interference is
destructive due to the imaginary nature of the Pomeron exchange amplitude at
$N_1$ and the factor of $i$ resulting from the intermediate state between
$N_1$ and $N_2$ going on-shell.

\begin{figure}[htb]
\begin{center}
\leavevmode
{\epsfxsize=14truecm \epsfbox{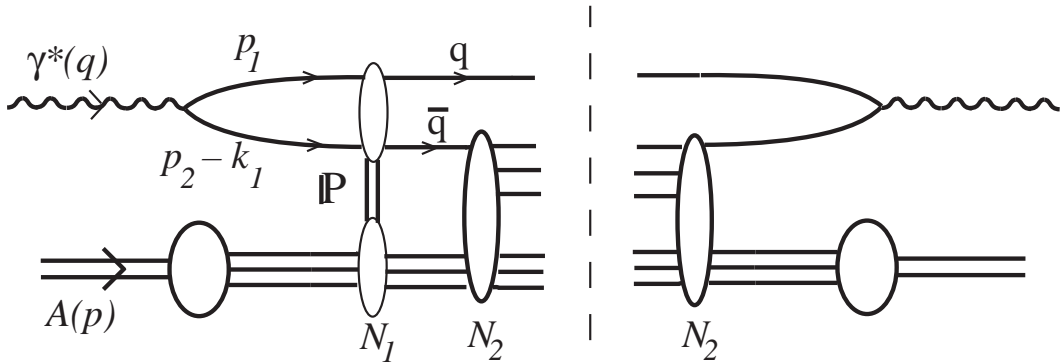}}
\end{center}
\caption[*]{Glauber-Gribov
shadowing involves interference between
rescattering amplitudes.}
\label{fig2dis}
\end{figure}

This shadowing effect on the DIS cross section is not compatible
with the cross section being determined by the parton probabilities ${\cal
P}$ of Eq. \eq{strfn}. Since the Pomeron amplitude in Fig. 2 is
imaginary it must involve on-shell intermediate states. But initial state
interactions in the target {\em before} the virtual photon is absorbed cannot
create on-shell intermediate states  -- they would constitute decay channels of
the target. We conclude that Glauber-Gribov shadowing involves final state
interactions and hence must be associated with the path ordered exponential in
\eq{melm}.

\vspace{.5cm}

{\bf 5. A Perturbative Example of Shadowing}

We shall construct a perturbative example of the physics of
Glauber-Gribov shadowing, which is simple enough to allow explicit
expressions for the scattering amplitudes at small $x_B$. We use
this example in section 6 to verify the general result of section
3 that final state interactions between target spectators do not
affect the DIS cross section, whereas rescattering of the struck
quark does.

In this section we use standard covariant perturbation theory in
Feynman gauge of a scalar abelian gauge theory.

We consider the forward $\gamma^* T \to \gamma^* T$ amplitude of
Fig.~3, the discontinuity of which gives a contribution to
$\sigma(\gamma^* T)$ at order $\alpha\alpha_s^4$ via the optical
theorem \eq{optical}. Since we may assume the charges of the
target $T$ and the `quark' \qu\ to be distinct, we can focus on
the gauge invariant set of diagrams in which the gluons are
exchanged between the quark pair and the target. Each gluon $k_i$
can couple to either the \qu\ or the \qb\ line, and all distinct
permutations of the gluon vertices are included.

\begin{figure}[htb]
\begin{center}
\leavevmode
{\epsfxsize=13.5truecm \epsfbox{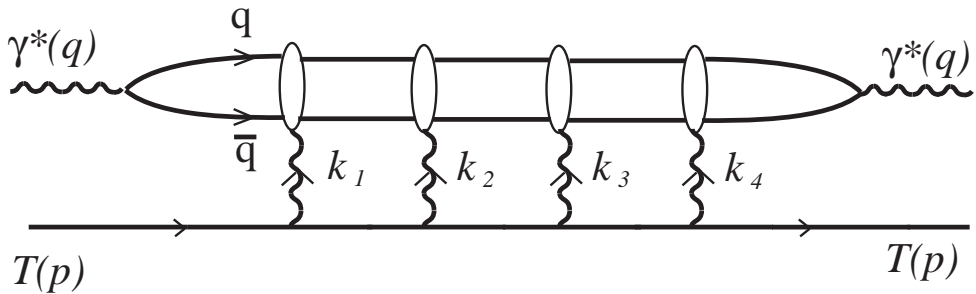}}
\end{center}
\caption[*]{Forward $\gamma^* T \to \gamma^* T$ amplitude. All
attachments of  the exchanged gluons to the upper scalar loop are included,
as well as  topologically distinct permutations of the lower vertices on the
target line.}
\label{fig3dis}
\end{figure}
Taking the discontinuity between gluons $k_3$ and $k_4$ gives a
contribution which models the interference term of Fig.~2. The
scattering on $N_2$ is given by single gluon exchange, while the
Pomeron exchange on $N_1$ is modelled by two gluon exchange. The
discontinuity between gluons $k_2$ and $k_3$ gives the square of
the `Pomeron' exchange amplitude. We calculate the one-, two- and
three-gluon exchange amplitudes for $\gamma^*T \to \qu\qbm T$
explicitly for $x_B \ll 1$, making use of the results of Ref.
\cite{bhm} where a similar model was studied. Since the target $T$
is taken to be elementary this model does not have shadowing in
the conventional sense described in section 4. It nevertheless
demonstrates how final state interference effects reduce the DIS
cross section.

We work in the target rest frame \eq{itrframe} and in the aligned jet
kinematics of Eqs. \eq{alkin} and \eq{p2plus}. The Feynman gauge calculation
is simplified by assuming\footnote{The expressions for the scattering
amplitudes that we derive at large $M$ are actually valid also when
$M$ and $k_{\perp}$ are of the same order. This is seen directly for the Born
amplitude of Fig.~4, and from the LC gauge calculations in the Appendix for the
loop amplitudes.} a large target mass $M$. Hence the kinematic limit
we consider
is
\beq
2\nu \sim p_1^- \gg M \gg p_2^- \gg k_{i\perp},\ p_{2\perp},\ k_i^-,
\ m \gg k_i^+,\ k^+ = Mx_B+ p_2^+
\label{Mscales}
\eeq
where $m$ is the mass of the \qu, \qb\ quarks and $k=\sum_i k_i$ is the
total momentum transfer from the target.

\vspace{.25cm}

{\it 5.1 Single Gluon Exchange Amplitude $A$}

The three Feynman diagrams are shown in Fig.~4. As in section 3 we
use the virtual photon polarization vectors \eq{polvec} and find
that the dominant (leading twist) contribution comes from
$\varepsilon(\lambda=0)\cdot p_1 \simeq Q$. Diagram 4c is
proportional to $\varepsilon\cdot (p+p')$ and is thus non-leading.
Diagram 4a involves the quark propagator \beq \label{p2kprop}
(p_2-k)^2 - m^2 \simeq p_2^- (p_2^+ - k^+)
-(\pvec_{2\perp}-\kvec_\perp)^2 -m^2 = - D(\pvec_{2\perp} -
\kvec_{\perp}) \eeq where we used \eq{kexch} and defined \beq
D(\pvec_{\perp}) \equiv p_2^- Mx_B + {p_\perp}^2 + m^2
\label{Ddef} \eeq Similarly the quark propagator in diagram 4b
gives $D(\pvec_{2\perp})$. The full amplitude in the limit
\eq{Mscales} is \beq A(p_2^-,\pvec_{2\perp},\kvec_\perp) =
\frac{2eg^2 M Q p_2^-}{k_\perp^2}\left[\frac{1}{D(\pvec_{2\perp})}
- \frac{1}{D(\pvec_{2\perp}-\kvec_\perp)} \right]  \label{Aexpr}
\eeq

\begin{figure}[htb]
\begin{center}
\leavevmode
{\epsfxsize=13.5truecm \epsfbox{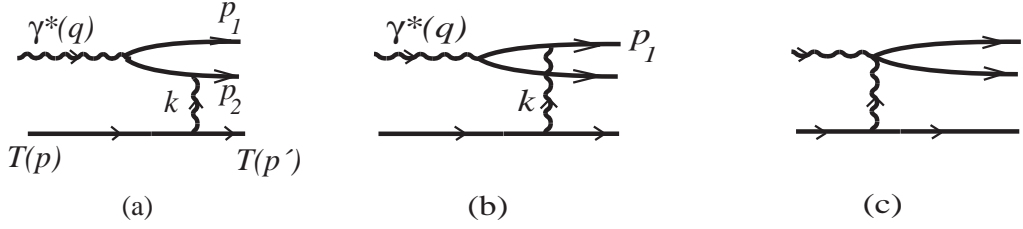}}
\end{center}
\caption[*]{Single gluon exchange diagrams in scalar abelian theory.}
\label{fig4dis}
\end{figure}

We may readily verify that this contribution is of leading twist. The
$\ell + T \to \ell' + X$ DIS cross section is\footnote{Here the lepton $\ell$
is assumed to have spin $\frac{1}{2}$.},
\beq
Q^4\frac{d\sigma}{dQ^2\, dx_B} = \frac{\alpha}{16\pi^2}\frac{1-y}{y^2}
\frac{1}{2M\nu} \int \frac{dp_2^-}{p_2^-}\,
\frac{d^2\pvec_{2\perp}}{(2\pi)^2}\, \frac{d^2\kvec_\perp}{(2\pi)^2}\, |A|^2
\label{discross}
\eeq
where $y=\nu/E_\ell$. The factor $Q^2$ in $|A|^2$ combines with $1/2M\nu$
in \eq{discross} to make the rhs independent of $Q^2$ in the Bjorken limit,
when the soft momenta $\kvec_\perp$ and $p_2$ are integrated over any finite
domain.

We also note that the dominant contribution to the DIS cross section at small
$x_B$ comes from $p_2^+ \sim Mx_B$ and $p_2^- \sim (p_{2\perp}^2+m^2)/Mx_B$ as
assumed in \eq{p2plus}. To see this, note that the amplitude $A \propto p_2^-$
for $p_2^- \ll (p_{2\perp}^2+m^2)/Mx_B$, while $A \propto 1/p_2^-$ for
$p_2^- \gg
(p_{2\perp}^2+m^2)/Mx_B$.

Since $A \propto 1/k_\perp$ for $k_\perp \to 0$ the cross section
\eq{discross} has a logarithmic singularity in this limit, which is regulated
by the longitudinal momentum exchange at $k_\perp \sim k^+ \sim Mx_B$. This
logarithmic behavior occurs only at lowest order \cite{BetheMaximon} and
will not be relevant for our conclusions.

It is instructive to express the cross section also as an integral over the
transverse distances $r_\perp, R_\perp$ conjugate to $p_{2\perp}, k_\perp$.
Defining
\beq
\tilde A(p_2^-,\rvec_\perp, \Rvec_\perp) = \int
\frac{d^2\pvec_{2\perp}}{(2\pi)^2}\, \frac{d^2\kvec_\perp}{(2\pi)^2}\,
A(p_2^-,\pvec_{2\perp},\kvec_\perp) \exp\left(i\rvec_\perp\cdot\pvec_{2\perp}
+ i\Rvec_\perp\cdot\kvec_{\perp} \right)
\label{Atilde}
\eeq
and using
\beq
V(m\, r_\perp) \equiv
\int \frac{d^2\pvec_\perp}{(2\pi)^2}
\frac{e^{i\rvec_\perp\cdot\pvec_{\perp}}}{p_\perp^2+m^2}
= \frac{1}{2\pi}K_0(m\,r_\perp)
\label{Vexpr}
\eeq
where $K_0$ is a Bessel function, and
\beq
W(\rvec_\perp, \Rvec_\perp) \equiv
\int \frac{d^2\kvec_\perp}{(2\pi)^2}
\frac{1-e^{i\rvec_\perp\cdot\kvec_{\perp}}}{k_\perp^2}
e^{i\Rvec_\perp\cdot\kvec_{\perp}} = \frac{1}{2\pi}
\log\left(\frac{|\Rvec_\perp+\rvec_\perp|}{R_\perp} \right)
\label{Wexpr}
\eeq
we get from \eq{Aexpr},
\beq
\tilde A(p_2^-,\rvec_\perp, \Rvec_\perp) = 2eg^2 M Q p_2^-\,
V(m_\pl r_\perp) W(\rvec_\perp, \Rvec_\perp)
\label{Atildeexpr}
\eeq
where
\beq
m_\pl^2 = p_2^-Mx_B + m^2 \label{mplus}
\eeq
The contribution \eq{discross} to the DIS cross section can then be expressed
as
\beqa
Q^4\frac{d\sigma}{dQ^2\, dx_B} &=&
\frac{\alpha^2\alpha_s^2}{\pi^3}\frac{1-y}{y^2} x_B M^2\int dp_2^-
\frac{p_2^-}{\left(p_2^-Mx_B + m^2 \right)^2}
\nonumber\\
&\times& \int d^2\vec u_\perp
d^2\vec U_\perp \left[K_0(u_\perp) \log\left(\frac{|\vec U_\perp+\vec
u_\perp|}{U_\perp}\right) \right]^2
\label{fourcross}
\eeqa
Here the dimensionless integration variables were defined as $\vec u_\perp =
\rvec_\perp m_\pl$ and $\vec U_\perp = \Rvec_\perp m_\pl$, showing that the
typical transverse distances $\rvec_\perp, \Rvec_\perp$ scale as $1/m_\pl$. The
$p_2^-$ integral in \eq{fourcross} is logarithmic\footnote{
We also note that \eq{fourcross} contains a collinear singularity
when $m \rightarrow 0$. In this limit the exchanged gluon becomes
a {\em collinear line} in the language of Ref. \cite{css}.}
at large $ p_2^- > m^2/Mx_B$, where the aligned jet $\gamma^* \qu \to \qu$
subprocess turns into $\gamma^*\gamma \to \qu\qbm$ \cite{bhm}.

\vspace{.25cm}

{\it 5.2 Two-Gluon Exchange Amplitude $B$}
Fig.~5 shows two of the
altogether six two-gluon exchange diagrams which give leading
contributions to the $\gamma^* T \to \qu\qbm T$ amplitude for $x_B
\ll 1$ in Feynman gauge. Diagrams with 4-point vertices (\cf\
Fig.~4c) are again suppressed in this gauge. We illustrate the
calculation of this one-loop amplitude using the diagrams of
Fig.~5.

\begin{figure}[htb]
\begin{center}
\leavevmode
{\epsfxsize=13.5truecm \epsfbox{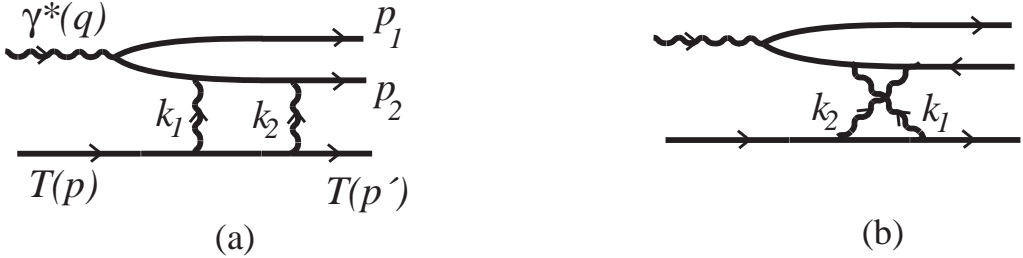}}
\end{center}
\caption[*]{Double gluon exchange diagrams. In Feynman gauge four more
diagrams contribute at leading order, where one or both of the exchanged
gluons attach to the quark ($p_1$) line.}
\label{fig5dis}
\end{figure}

Our assumption \eq{Mscales} of a large target mass $M$ simplifies the loop
integral by suppressing the $k_i^0$ momentum components. For the overall
exchange we find from the mass-shell condition $p'^2 = (p-k)^2 = M^2$ that
\beq
k^0 = k_1^0 + k_2^0 = -\frac{k_\perp^2}{2M} \ll k^\pm,\ k_\perp  \label{kzero}
\eeq
The corresponding suppression for the loop momentum $k_1^0 \simeq -k_2^0$
results from the sum of the uncrossed and crossed gluon attachments to the
target
line in Fig. 5,
\beqa
(-ig\,2M)^2\left[ \frac{i}{(p-k_1)^2-M^2+\ieps} +
\frac{i}{(p-k_2)^2-M^2+\ieps}
\right] && \nonumber\\
&&\nonumber\\
\simeq 2ig^2 M \left( \frac{1}{k_1^0-\ieps} + \frac{1}{k_2^0-\ieps} \right)
\simeq 2ig^2 M\ 2\pi i \delta(k_1^0) &&
\label{l2}
\eeqa

Making use of Eqs. \eq{p2kprop} and \eq{l2} we find
\beqa \label{5abexpr}
B(5a)+B(5b) &=& -\frac{2eg^4 MQ\,p_2^-}{D(\pvec_{2\perp}-\kvec_\perp)}
\int\frac{d^2 \kvec_{2\perp}\,dk_2^+}{(2\pi)^3}\frac{1}{\kvec_1^2
\kvec_2^2} \nonumber\\
&\times& \frac{1}{k_2^+ -(2\pvec_{2\perp}\cdot\kvec_{2\perp}
-\kvec_2^2)/p_2^- -\ieps}
\eeqa
Symmetrizing the integrand in $\kvec_1 \leftrightarrow \kvec_2$ and recalling
\eq{kexch} the last factor becomes
\beqa \label{k2plusconstr}
\frac{1}{2}\left[\frac{1}{k_2^+ -(2\pvec_{2\perp}\cdot\kvec_{2\perp}
-\kvec_2^2)/p_2^- -\ieps} -\frac{1}{k_2^+ -k^+
+(2\pvec_{2\perp}\cdot\kvec_{1\perp} -\kvec_1^2)/p_2^- +\ieps} \right]&&
\nonumber\\&&\\
\simeq i\pi\delta(k_2^+) \mbox{\hspace{10cm}}&& \nonumber
\eeqa
Thus
\beq \label{5abresult}
B(5a)+B(5b) = -\frac{ieg^4MQ\,p_2^-}{D(\pvec_{2\perp}-\kvec_\perp)}
\int\frac{d^2 \kvec_{2\perp}}{(2\pi)^2}\,\frac{1}{k_{1\perp}^2\,k_{2\perp}^2}
\eeq

Adding the contributions from the remaining four diagrams we find for the full
two-gluon exchange amplitude
\beqa
B(p_2^-,\pvec_{2\perp},\kvec_\perp) &=& -ieg^4 M Q p_2^- \int \frac{d^2
\kvec_{1\perp}}{(2\pi)^2}\ \frac{1}{k_{1\perp}^2 k_{2\perp}^2}
\label{Bexpr}\\
&\times& \left[
\frac{1}{D(\pvec_{2\perp})} - \frac{1}{D(\pvec_{2\perp}-\kvec_{1\perp})}
-\frac{1}{D(\pvec_{2\perp}-\kvec_{2\perp})}+
\frac{1}{D(\pvec_{2\perp}-\kvec_{\perp})} \right] \nonumber
\eeqa
where $\kvec_{2\perp}=\kvec_{\perp} -\kvec_{1\perp}$. We note that
the amplitude
is fully imaginary as required by crossing symmetry, since $B \propto p_2^-$ as
$p_2^- \to \infty$ and two gluon exchange has even charge conjugation. Thus our
model captures the essential features of Pomeron exchange. We note also that
$B(p_2^-,\pvec_{2\perp},\kvec_\perp) \propto \log k_\perp$ for $k_\perp \to 0$.
In contrast to the single gluon exchange contribution to the DIS cross section,
the square of \eq{Bexpr} can thus be safely integrated over $k_\perp$ and
(for $m
\neq 0$) over $p_2^-$.

Due to conservation of the transverse distances $\rvec_\perp,
\Rvec_\perp$ in the peripheral scattering, the Fourier transform
\eq{Atilde} returns the simple form \beq \tilde
B(p_2^-,\rvec_\perp, \Rvec_\perp) = -ieg^4 M Q p_2^- V(m_\pl
r_\perp) W^2(\rvec_\perp, \Rvec_\perp)= \frac{-ig^2}{2!} W \tilde
A \label{Btildeexpr} \eeq where we used \eq{Vexpr} and \eq{Wexpr}.

We stress that in the $x_B \to 0$ limit, the amplitude $B$ is
dominated by the configuration where the intermediate state
between the two exchanges is on-shell. This can be seen by
calculating $B$ in LC time-ordered perturbation theory, where this
intermediate state is associated with a vanishing denominator
\eq{dint}. Alternatively, we may note that since the real part of
$B$ is suppressed in the $x_B \to 0$ limit the full amplitude is
(via the optical theorem) given by its discontinuity. This is true
in all gauges since $B$ is gauge invariant.

\vspace{.25cm}

{\it 5.3 Three-Gluon Exchange Amplitude $C$}

No qualitatively new aspects appear
in the calculation of this two-loop amplitude. Permuting the attachments of
the three gluons on the target line one finds in analogy to \eq{l2} that
$k_i^0 \simeq 0$ for all exchanges $(i=1,2,3)$. Similarly the $k_i^+$
integrations are simply evaluated after symmetrizations analogous to
\eq{k2plusconstr}. The final expression in momentum space is
\beqa
C(p_2^-,\pvec_{2\perp},\kvec_\perp) = -\frac{1}{3} eg^6 M Q p_2^-\int
\frac{d^2 \kvec_{1\perp}}{(2\pi)^2}
\frac{d^2 \kvec_{2\perp}}{(2\pi)^2}\ \frac{1}{\kvec_{1\perp}^2\,
\kvec_{2\perp}^2\, \kvec_{3\perp}^2} \mbox{\hspace{3.5cm}} \label{Cexpr} &&
\\&&\nonumber\\
\times \left[ \frac{1}{D(\pvec_{2\perp})} -\frac{3}{D(\pvec_{2\perp}
-\kvec_{1\perp})} +
\frac{3}{D(\pvec_{2\perp}- \kvec_{1\perp}-\kvec_{2\perp})} -
\frac{1}{D(\pvec_{2\perp} -\kvec_{\perp})} \right] && \nonumber
\eeqa
where $\kvec_{3\perp}=\kvec_{\perp}- \kvec_{1\perp}- \kvec_{2\perp}$.

The Fourier transform \eq{Atilde} gives the amplitude in transverse
coordinate space as
\beq
\tilde C(p_2^-,\rvec_\perp, \Rvec_\perp) = -\frac{1}{3}eg^6 M Q p_2^-
V(m_\pl r_\perp) W^3(\rvec_\perp, \Rvec_\perp)=
\frac{(-ig^2)^2}{3!}W^2 \tilde A
\label{Ctildeexpr}
\eeq

Similarly to the $B$ amplitude, $C$ arises from the intermediate
states between the rescatterings being on-shell in the $x_B \to 0$
limit. Again this must hold also in LC gauge. Since in the two-loop
case there are two consequtive intermediate states, $C$ is purely real.

From the expressions \eq{Atildeexpr}, \eq{Btildeexpr} and
\eq{Ctildeexpr}, it is apparent that the sum of gluon-exchange
amplitudes exponentiates, \beq \tilde{\cal M}(p_2^-,\rvec_\perp,
\Rvec_\perp) = \tilde A + \tilde B + \tilde C\ldots = -2ieM Q
p_2^- V \left[1-\exp(-ig^2 W) \right] \label{Mtildeexpr} \eeq

As noted at the beginning of this section, we have assumed the
charges of the quark and target lines to be distinct. This allows
us to restrict our analysis to the subclass of Feynman diagrams
considered above, since diagrams with different powers of the
charges cannot cancel in the DIS cross section. However, we should
note that at the level of three gluon exchanges there are new
types of diagrams which have the same charge dependence as $C$ in
Eq. \eq{Cexpr}. For example, one of the three gluons may be
exchanged between the quarks while another forms a loop on the
target line. The $k_\perp$-dependence of this contribution would
differ from that of \eq{Cexpr}. We do not further consider such
contributions.

\vspace{.5cm}

{\bf 6. Effects of Rescattering on the DIS Cross Section}

We now use our perturbative amplitudes to demonstrate that final-state
rescattering of the struck quark affects the DIS cross section. In the previous
section we used covariant (rather than time-ordered) perturbation theory, and
thus did not distinguish between initial (ISI) and final (FSI) state
interactions. However, diagrams involving rescattering of the struck quark
necessarily are FSI because the exchanged gluon couples to the struck quark
($p_1$) line {\em after} the virtual photon.  We shall see that precisely such
diagrams contribute to the cross section.

We consider the DIS cross section \eq{discross} expressed as a sum over the
transverse distances $\rvec_\perp, \Rvec_\perp$ defined in \eq{Atilde},
\beq
Q^4\frac{d\sigma}{dQ^2\, dx_B} = \frac{\alpha}{16\pi^2}\frac{1-y}{y^2}
\frac{1}{2M\nu} \int \frac{dp_2^-}{p_2^-}\,d^2\rvec_\perp\, d^2\Rvec_\perp\,
|\tilde\M|^2
\label{transcross}
\eeq
where
\beq
|\tilde{\cal M}(p_2^-,\rvec_\perp,
\Rvec_\perp)| =
\left|\frac{\sin \left[g^2\, W(\rvec_\perp,
\Rvec_\perp)/2\right]}{g^2\, W(\rvec_\perp,
\Rvec_\perp)/2} \tilde{A}(p_2^-,\rvec_\perp, \Rvec_\perp)\right|
\label{Interference}
\eeq
is the resummed amplitude \eq{Mtildeexpr} and $V, W$ are given in Eqs.
\eq{Vexpr} and \eq{Wexpr}, respectively.

The fact that the coefficient of $\tilde A$ in \eq{Interference}
is less than unity for all $\rvec_\perp, \Rvec_\perp$ shows that
the rescattering corrections included in $\tilde\M$ reduce the
cross section. This effect agrees with the Glauber-Gribov picture
of DIS shadowing and must be present also in LC gauge (see section
  7).

The forward $\gamma^* T \to \gamma^* T$ amplitude in Fig.~3 can also be cut
through some of the gluon lines, corresponding to final states with real
gluons. Such contributions have, however, a different target mass $M$
dependence (\cf\ Eq. \eq{kzero}). Similar arguments suggest that other
contributions, even if they are of the same order in the coupling constants,
cannot change the conclusion that the DIS cross section is influenced by final
state interactions.

In section 3 we gave a general argument (in Feynman gauge) which
showed that final state interactions between target spectators
cannot influence the DIS cross section (\cf\ Fig. 1a). We shall
now check this statement using our perturbative amplitudes.

In the aligned jet kinematics the antiquark belongs to the target system. We
thus consider the subset of diagrams like Figs.~4a and 5 where all exchanged
gluons attach to the \qb\ ($p_2$) line. One can easily verify that this subset
of  diagrams is gauge invariant in the class of covariant gauges in our
kinematic
limit \eq{Mscales}. The corresponding sum of cuts in Fig.~3 is then
proportional
to
\beq
S_{\qbm}(p_2^-,\pvec_{2\perp}, \kvec_\perp) =
    |B_{\qbm}|^2 + 2{\rm Re} (A_{\qbm}\, C_{\qbm}^*) \label{sqbar}
\eeq
where the subscript \qb\ indicates the subset of diagrams.

Diagrams where all gluons attach to the antiquark line can involve both ISI and
FSI. Since the two-gluon exchange contribution \eq{5abresult} is imaginary it
must, however, involve rescattering of on-shell intermediate states which can
only arise after the virtual photon has been absorbed. Similarly the (real)
three-gluon exchange amplitude $C$ \eq{Cexpr} involves double rescattering of
on-shell states. Hence all our amplitudes (except the Born term $A$)
involve FSI.

It is straightforward to identify the $A_{\qbm}, B_{\qbm}, C_{\qbm}$
contributions to the expressions \eq{Aexpr}, \eq{Bexpr}, \eq{Cexpr} of the
full one-, two- and three-gluon exchange amplitudes in momentum
space. According
to Eq. \eq{p2kprop} the antiquark propagator next to the virtual photon vertex
gives a denominator $D(\pvec_{2\perp}-\kvec_{\perp})$ for all diagrams in our
subset. This factor appears explicitly in each amplitude. Dimensionally
regularizing the logarithmic infrared divergencies at $k_{i\perp}=0$ we
thus find
\beq
S_{\qbm}(p_2^-,\pvec_{2\perp}, \kvec_\perp) =
    \left[\frac{eg^4 MQp_2^-}{(2\pi)^2
D(\pvec_{2\perp}-\kvec_{\perp})} \right]^2
\left\{
    \left[R_2(k_{\perp})\right]^2 -\frac{4}{3} R_{13}(k_{\perp})  \right\}
\label{sqbarbis}
\eeq
where
\beqa
R_2(k_{\perp}) &=& \int \frac{d^D\kvec_{1\perp}}{k_{1\perp}^2(\kvec_\perp
-\kvec_{1\perp})^2}
=\frac{\pi^{D/2}}{k_{\perp}^{4-D}}
\frac{\left[\Gamma\left(\frac{D}{2}-1\right)\right]^2
\Gamma\left(2-\frac{D}{2}\right)}{\Gamma\left(D-2\right)}
    \nonumber \\ && \label{Rdef} \\
R_{13}(k_{\perp}) &=& \frac{1}{k_\perp^2} \int
\frac{d^D\kvec_{1\perp}\ d^D\kvec_{2\perp}}{k_{1\perp}^2
k_{2\perp}^2(\kvec_\perp -\kvec_{1\perp} -\kvec_{2\perp})^2}
=\frac{\pi^{D}}{k_{\perp}^{8-2D}}
\frac{\left[\Gamma\left(\frac{D}{2}-1\right)\right]^3
\Gamma\left(3-D\right)}{\Gamma\left(\frac{3D}{2}-3
\right)}
\nonumber
\eeqa

Expanding $R_2$ and $R_{13}$ around $D=2$ gives
\begin{eqnarray}
R_2(k_{\perp})&=&\frac{\pi^{D/2}}{k_{\perp}^{4-D}}
\left\{\frac{4}{D-2}+2\gamma+ \frac{1}{12}\left(6\gamma^2 -
\pi^2\right)\left(D-2\right)
+\right. \nonumber \\ &+& \left.
\frac{1}{24}\left[2\gamma^3 -\gamma \pi^2-14
\psi^{(2)}(1)\right]\left(D-2\right)^2 + {\cal
O}\left[(D-2)^3\right] \right\} \ ,
\nonumber\\&& \label{Rexpr} \\
R_{13}(k_{\perp})
&=&\frac{\pi^{D}}{k_{\perp}^{8-2D}}\left\{\frac{12}{\left(D-2\right)^2}+
\frac{12\gamma}{D-2}+\left(6\gamma^2 - \frac{\pi^2}{2}\right)
+\right. \nonumber \\ &+& \left.
\frac{1}{2} \left[4\gamma^3 - \gamma\pi^2
    - 16\psi^{(2)}(1)\right]\left(D-2\right) + {\cal
O}\left[(D-2)^2\right]\right\} \ ,  \nonumber
\end{eqnarray}
where $\gamma\simeq 0.577$ is Euler's constant and $\psi^{(n)}(z)$
is the $(n+1)^{\rm th}$ logarithmic derivative of the gamma
function with $\psi^{(2)}(1)\simeq -2.40$. Hence \beqa
\left[R_2(k_{\perp})\right]^2 -\frac{4}{3} R_{13}(k_{\perp}) =
\frac{\pi^D}{k_\perp^{8-2D}} \, 6 \psi^{(2)}(1) \left(D-2\right) +
{\cal O}\left[(D-2)^2\right]  \label{Dexp} \eeqa vanishes at
$D=2$. Thus FSI between the target spectators do not change the
DIS cross section.

We conclude that in covariant gauges, only final state
interactions which involve rescatterings of the current quark
affect the DIS cross section.

\vspace{.5cm}

{\bf 7. Light-Cone Gauge $A^+=0$}

We have seen that in a covariant gauge, the DIS cross section is
influenced by final-state interactions of the struck quark in the
target. This soft physics is contained in the path-ordered
exponential of the matrix element \eq{melm} in a general gauge and
appears to vanish in LC gauge, $n\cdot A=A^+=0$.

However, as we have seen in section 5 the amplitudes $B$ and $C$
arise from on-shell intermediate states in the $x_B \to 0$ limit.
Thus in \eq{transcross} the contribution of $|\tilde\M|^2$, whose
expansion starts as $|\tilde B|^2 + 2 \tilde A \tilde C$, also
arises purely from on-shell intermediate states. The presence of
such on-shell states is gauge independent and they can only occur
in the final state. We conclude that the DIS cross section is
influenced by final state interactions in all gauges. Thus parton
distributions cannot be fully determined by parton probabilities
in the target.

Let us now discuss some features inherent to the LC gauge
preventing parton distributions from being probabilities, in other
words making the expression \eq{melm} for $f_{\qu/N}$ incorrect in
LC gauge. It turns out that terms which are next-to-leading
corrections in a general gauge cannot be ignored in LC gauge. To
see this, it is helpful to recall how the exponential arises from
perturbative diagrams.

As explained in Ref. \cite{css} each quark field is associated with an
ordered exponential
\beq \label{poe}
[A^+] \equiv {\rm P}\exp\left[ig\int_0^{\infty} dy^- A^+(y^-) \right]
\eeq
where the gauge field $A^+$ is evaluated on the light-cone, $y^+=
y_\perp=0$. This factor arises from the interactions of the struck quark as
it moves through the target. While the path in \eq{poe} extends to infinity,
there is a partial cancellation between the quark fields in the matrix
element \eq{melm} leaving a path of length $y^-\sim 1/Mx_B$ equal to the
coherence length of the virtual photon. Only interactions within this LC
distance can influence the cross section.

Expanding the exponential \eq{poe} gives
\beqa \label{pathexpand}
[A^+]&=& 1+ig\int_0^\infty dy_1^- A^+(y_1^-) \left[ 1 + ig \int_{y_1^-}^\infty
dy_2^- A^+(y_2^-) + \ldots \right] \nonumber\\
&& \\
&=& 1+g\int_{-\infty}^\infty \frac{dk_1^+}{2\pi} \frac{\tilde A^+(k_1^+)}
{k_1^+-\ieps}+g^2\int_{-\infty}^\infty \frac{dk_1^+ dk_2^+}{(2\pi)^2}
\frac{\tilde A^+(k_1^+)\tilde A^+(k_2^+)} {(k_1^+ +k_2^+-\ieps)
(k_2^+-\ieps)} + \ldots \nonumber
\eeqa
where
\beq \label{fta}
A^+(y^-) = \int_{-\infty}^\infty \frac{dk^+}{2\pi} \tilde A^+(k^+)
\exp(-ik^+ y^-)
\eeq

\begin{figure}[htb]
\begin{center}
\leavevmode
{\epsfxsize=13.5truecm \epsfbox{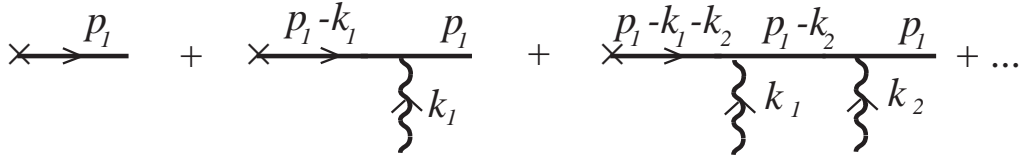}}
\end{center}
\caption[*]{Scattering of the struck quark on the gauge field of the target
which gives rise to the ordered exponential \eq{poe}.}
\label{fig6dis}
\end{figure}

The terms in the expansion \eq{pathexpand} arise from the perturbative
diagrams of Fig. 6, where the cross indicates the virtual photon vertex. The
struck quark momentum is asymptotically large, $p_1^- \to\infty$, implying that
the quark moves along the light-cone $y^+=y_\perp=0$. The two-gluon
exchange term
in Fig. 6 is given by
\beq \label{2int}
(ig)^2 i^2 \frac{p_1^- \tilde A^+(k_1^+)\,p_1^- \tilde A^+(k_2^+)}
{\left[-p_1^-(k_1^+ +k_2^+)+\ieps \right](-p_1^-k_2^+ +\ieps)} =
g^2\frac{\tilde A^+(k_1^+)\tilde A^+(k_2^+)}{(k_1^+ +k_2^+ -\ieps)(k_2^+
-\ieps)}
\eeq
Thus we find equivalence to the expression \eq{pathexpand} by approximating
$(2p_1-k_2)\cdot\tilde A(k_2^+) \simeq p_1^- \tilde A^+(k_2^+)$, \ie,
by keeping
only the asymptotically large component of $p_1$. This is correct in all gauges
{\em except} $A^+=0$, where this `leading' term actually vanishes.

Neglecting the dependence of the matrix element \eq{melm} on the gauge field
$\tilde A(k^+)$ in LC gauge is equivalent to assuming that interactions of
the struck quark with the gauge field such as $(2\pvec_1 -\kvec_2)_\perp
\cdot \vec{{\tilde A}}_\perp$ do not contribute at leading twist. The
following example shows how this assumption can fail.

As a simple illustration of how the high energy and LC gauge limit can fail
to commute we consider
the elastic process $\qu(p_1-k)\, T(p) \to \qu(p_1)\, T(p-k)$,
where $p=(M,M,\vec 0_\perp)$ and $p_1^- \to \infty$ at fixed $p_{1\perp},
k_\perp$. Momentum conservation implies
\beqa \label{kelastic}
k^+ &=& \frac{(2\pvec_{1\perp}-\kvec_\perp)\cdot \kvec_\perp}{p_1^-} \to 0
\nonumber\\
k^- &=& -\frac{k_\perp^2}{M}\ \ \ \mbox{fixed}
\eeqa
The interaction of the gauge field with the quark is given by
$(-ig)(2p_1-k)_\mu \cdot d^{\mu\nu}(k)$. In Feynman gauge the propagator is
$d_F^{\mu\nu}(k)=-ig^{\mu\nu}/(k^2+\ieps)$ and the coupling is dominated by
$-igp_1^-\,d_F^{+-}(k)$, which is analogous to the interaction \eq{2int} in
the ordered exponential. The elastic amplitude
\beq \label{elamp}
A(\qu T\to \qu T) = -2g^2M\frac{p_1^-}{k_\perp^2}
\eeq
is thus $\propto p_1^-$ as befits Coulomb exchange.

In LC gauge the propagator \eq{lcprop} satisfies $d_{LC}^{+\nu}(k)=0$,
hence the $p_1^-$ component does not contribute. Yet the elastic amplitude is
gauge independent and must still be given by \eq{elamp}. The absence of the
factor $p_1^-$ in the numerator coupling is in fact compensated by the factor
$k^+\propto 1/p_1^-$ in the denominator of the LC gauge propagator \eq{lcprop}.
The dominant contribution is from $-(2\pvec_1-\kvec)^\perp \cdot d^{-\perp}(k)$
and the result indeed agrees with \eq{elamp}.

Note that if we had kept only the $d^{+\nu}(k)$ part of the gauge
propagator in the high energy limit and then chosen LC gauge the
elastic scattering amplitude would have seemed to vanish. This
incorrect result is analogous to the apparent absence of
rescattering effects in the matrix element \eq{melm} for $A^+=0$.

In the Feynman gauge calculation of section 5 we saw that the
reinteractions of the struck quark with the target are essentially
elastic, the intermediate states being on-shell in the $x_B\to 0$
limit. It is thus not surprising that the calculation of the
scattering amplitudes in LC gauge has many features in common with
the elastic scattering example above. Details of the calculation
of the one-loop and two-loop amplitudes $B$ and $C$ \eq{Bexpr} and
\eq{Cexpr} in LC gauge are given in the Appendices.

In LC gauge the Feynman rules must be supplemented with a
prescription for the $k^+=0$ pole of the propagator \eq{lcprop}.
Three prescriptions that have been studied in the literature
\cite{lb,gl,kov} are given in Eq. \eq{prescriptions} of Appendix
A. The contributions of the individual diagrams shown in Fig. 7
for the one-loop amplitude $B$ depend on the prescription.
However, the $k_i^+=0$ poles cancel when all diagrams are added.
Their sum is thus prescription independent and agrees with the
Feynman gauge result \eq{Bexpr}. We verify the prescription
independence of the two-loop amplitude $C$ in Appendix B. A
consistent procedure for regulating the spurious poles is also
discussed there.

As we have already emphasized, final state interactions (FSI)
modify the DIS cross section also in LC gauge due to the presence
of on-shell intermediate states between the rescatterings in the
amplitudes $B$ and $C$. However, while in Feynman gauge it is the
rescattering of the struck quark $p_1$ which affects the cross
section, in LC gauge those rescatterings actually do not
contribute. Indeed, we show in Appendix C that contributions from
diagrams like Fig. 7c and 8b to the individual amplitudes cancel
in the cross section. Thus in LC gauge, independently of the
prescription, the cross section is modified by FSI occurring on
the antiquark $p_2$, \ie, within the target system. Choosing the
$A^+=0$ gauge shifts the rescatterings of the quark present in
Feynman gauge to rescatterings of the antiquark. As also shown in
Appendix C, in LC gauge the partial amplitude where only
attachments to $p_2$ are kept equals the full amplitude, up to a
phase factor. Which particle actually scatters in the
quark-antiquark system depends on the gauge, but the presence of
on-shell intermediate states does not.

Subtleties can appear when using the Kovchegov (K) prescription
(see Eq. \eq{prescriptions}), since the imaginary part arising
from a physical cut can be changed by the imaginary part created
by the prescription itself. The ${\rm K}$ prescription simulates
the physics of the rescattering corrections by introducing an
external gauge field into the dynamics. Unlike the Principal Value
(PV) or Mandelstam-Leibbrandt prescription, the K prescription is
not causal, and thus it would normally not be used for solving the
bound state problem and light-cone wave functions of an isolated
hadron in QCD. The solutions for the light-cone wave function of
the target hadron in the presence of an external gauge field can
have complex phases. This is apparently the way in which the
light-cone wave functions of a nucleus in the Kovchegov light-cone
gauge prescription mimic the effects of rescattering of the fast
quark and the Glauber-Gribov shadowing modifications of the
structure functions. If this picture could be validated, the
Kovchegov LC gauge prescription would give a framework in which
$\sigma_{DIS}$ is fully determined by the target LC wavefunction,
solved in the presence of an external field.

\vspace{.5cm}

{\bf 8. Conclusions}

We have found that final state Coulomb rescattering in the target,
within the coherence length $2\nu/Q^2 = 1/Mx_B$ of the hard process,
influences the $\ell N \to \ell' X$ DIS cross section.
In particular, diffractive scattering of the outgoing
quark-pair on target spectators is a coherent effect which is not included
in the light-front wave functions, even in light-cone gauge.
Such effects modify the contributions of the individual target partons,
implying that the DIS cross section is not fully given by the
parton probabilities of the initial state.
These coherent effects are reminiscent of the LPM effect
\cite{lpm}, which suppresses the bremsstrahlung of a high energy electron
in matter due to Coulomb rescattering of the electron within the formation
time of its radiated photons.

Our analysis, when interpreted in frames with $q^+ > 0,$ also
supports the color dipole description of deep inelastic lepton
scattering at small $x$.  Even in the case of the aligned jet
configurations, one can understand DIS as due to the coherent
color gauge interactions of the incoming quark-pair state of the
photon interacting first coherently and finally incoherently in
the target.

Our analysis in light-cone gauge resembles the ``covariant parton
model" of Landshoff, Polkinghorne and
Short~\cite{Landshoff:1971ff,Brodsky:1973hm} in the target rest
frame.  In this description of small $x$ DIS, the virtual photon
with positive $q^+$ first splits into the pair $p_1$ and $p_2$.
The aligned quark $p_1$ has no final state interactions.  However,
the antiquark line $p_2$ can interact in the target with an
effective energy $\hat s \propto {k_\perp^2/x}$ while staying
close to its mass shell.  Thus at small $x$ and large $\hat s$,
the antiquark $p_2$ line can first multiple scatter in the target
via pomeron and Reggeon exchange, and then finally scatter
inelastically or be annihilated. The DIS cross section can thus be
written as an integral of the $\sigma(\bar q p \to X)$ cross
section over the $p_2$ virtuality. In this way, the diffractive
scattering of the antiquark in the nucleus gives rise to the
shadowing of the nuclear cross section $\sigma(\bar q A \to X)$
\cite{Brodsky:1990qz}.

Our results do not contradict the QCD factorization theorem
\cite{css} for inclusive reactions in a general gauge. However,
they show that the apparent equivalence between the DIS cross
section and the target parton probabilities \eq{strfn} suggested
by the forward matrix element \eq{melm} in $A^+=0$ gauge is
incorrect. The $A_\perp$ components of the gauge field give
leading twist contributions in LC gauge.

Our investigation was triggered by the fact that the physically
plausible and phenomenologically successful Glauber-Gribov
description of DIS shadowing \cite{gribov,pw} implies that final
state interactions influence the DIS cross section. The physics of
shadowing is associated with final state diffractive scattering
rather than with the (real) light-cone wave function of the
target. There remains the possibility of incorporating shadowing
in the target wave function by solving it under the specific
boundary conditions implied by the Kovchegov LC gauge prescription
\cite{kov}.

Our analysis is consistent with the standard Operator Product
Expansion. Hence the usual sum rules of the parton distributions
remain valid in spite of the rescattering (shadowing) physics. We
have not estimated the quantitative importance of the rescattering
effects on $\sigma_{DIS}$, but it is natural to expect that they
are more prominent at small values of $x_B$ where the coherence
length is long. In particular, diffractive DIS is related to
shadowing and is apparently generated by rescattering
contributions.

\vspace{.5cm}

{\bf Acknowledgements.} We wish to thank G. Bodwin, T. Binoth, J.
D. Bjorken, M. Burkardt, J. Collins, L. Frankfurt, A. Hebecker, G.
Heinrich, Y. Kovchegov, L. Mankiewicz and M. Strikman for useful
discussions. SJB and PH are grateful for the hospitality and
support of the Institute for Nuclear Theory at the University of
Washington during the completion of this work.

\vspace{1cm}

\centerline{{\bf APPENDIX}}

\vspace{.5cm}

\centerline{{\bf A. One-loop calculation in $A^+=0$ gauge.}}

In this Appendix we present the calculation of the two-gluon exchange
amplitude $B$ \eq{Bexpr} in light-cone $n \cdot A = A^+=0$ gauge of a scalar
abelian theory. We shall take the target mass to be of the order of the
transverse momenta, \ie, rather than \eq{Mscales}, we here consider the
kinematic
limit
\beq
2\nu \sim p_1^- \gg p_2^- \gg k_{i\perp},\ p_{2\perp},\ k_i^-,\ m,\ M \gg
\ k_i^+,\ k^+=Mx_B + p_2^+
\label{scales}
\eeq
and show that the expression for the amplitude remains the same.

\begin{figure}[htb]
\begin{center}
\leavevmode
{\epsfxsize=13.5truecm \epsfbox{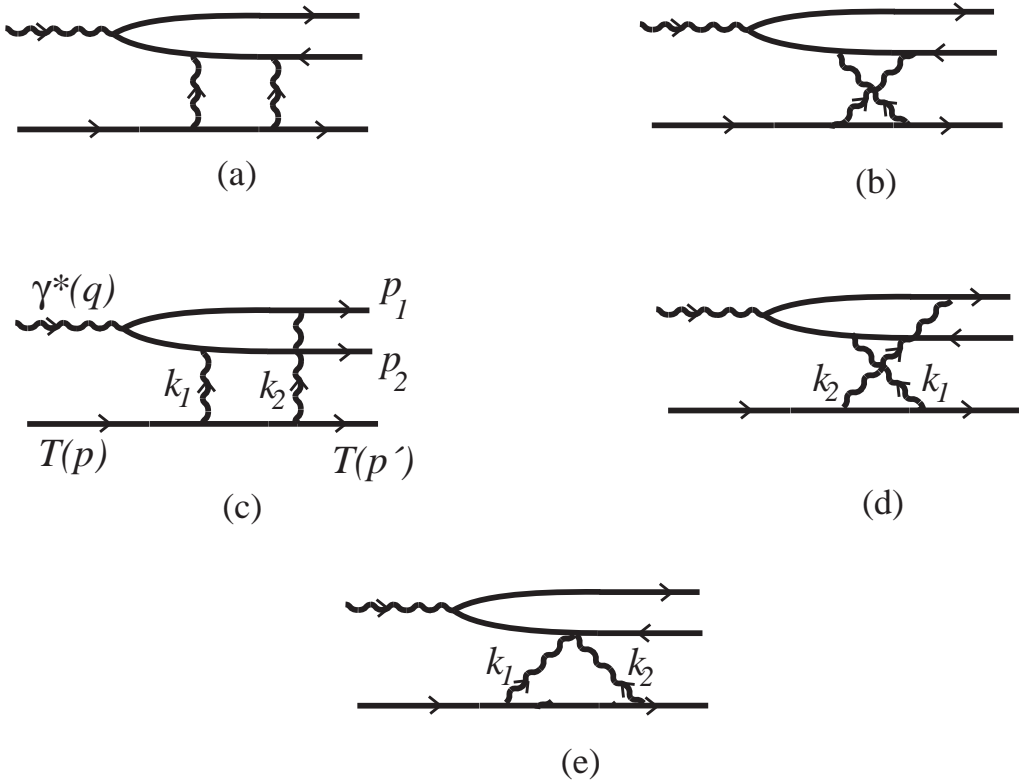}}
\end{center}
\caption[*]{Diagrams that can give leading order contributions to the
one-loop amplitude $B$ in $A^+=0$ gauge.}
\label{fig7dis}
\end{figure}

Leading contributions to the amplitude can come from diagrams $B_a \ldots
B_e$ of Fig. 7. The factors associated with the gluon propagators are
approximated as
\beq
(p+p')^{\mu} d_{\mu \nu}(k) (2l+k)^{\nu} \simeq i \frac{2M}{k_\perp^2 k^+}
\left[ D(k+l)-D(l) \right]
\label{A1}
\eeq
where only the $d^{- \perp}$ part of the propagator \eq{lcprop} contributes,
and the function $D(p) \equiv D(\pvec_{\perp})$ is defined in \eq{Ddef}.
Similarly, the factor from the four-leg scalar abelian vertex simplifies to
\beq
(2p-k_1)^{\mu} d_{\mu \nu}(k_1) d^{\mu' \nu}(k_2) (2p'+k_2)_{\mu'}
\simeq  \frac{(2M)^2}{k_1^+ k_2^+} \, \frac{\kvec_{1 \perp} \cdot \kvec_{2
\perp}}{k_{1\perp}^2 k_{2\perp}^2}
\label{A2}
\eeq
where again the $d^{- \perp}$ components dominate. A factor $2ig^2$ has
been omitted for the time being.  Direct use of the Feynman rules and of the
kinematics \eq{scales} leads to:
\beqa \label{A3}
B_a + B_b &=& -eg^4 QM \int_{\perp} \int \frac{dk_1^-}{2i\pi} \,
J(k_1^-) \int \frac{dk_2^+}{\pi}
   \frac{1}{k_1^+ k_2^+} \nonumber \\
&\ &\times \frac{[D(p_1)-D(p_2-k_2)][D(p_2-k_2)-D(p_2)]}{-D(p_1)
[-p_2^- k_2^+ + D(p_2)-D(p_2-k_2) +i\epsilon]} \nonumber \\
B_c + B_d &=& -eg^4 QM \int_{\perp} \int \frac{dk_1^-}{2i\pi} \,
J(k_1^-) \int \frac{dk_2^+}{\pi} \, \frac{1}{k_1^+ k_2^+} \nonumber \\
&\times& \frac{[D(p_1)-D(p_2-k_1)][D(p_2-k_1)-D(p_2)]}{
[-p_1^- k_2^+ + D(p_1)-D(p_2-k_1) +i\epsilon]
[p_2^- k_2^+ -D(p_2-k_1) +i\epsilon]} \nonumber \\
    B_e &=& -eg^4 QM \int_{\perp}  \int \frac{dk_1^-}{2i\pi} \,
J(k_1^-) \int \frac{dk_2^+}{\pi} \, \frac{1}{k_1^+ k_2^+} \,
\frac{\kvec_{1 \perp} \cdot \kvec_{2 \perp}}{-D(p_1)}
\eeqa
where we use the shorthand notations
\beqa
\int_{\perp} &\equiv& \int \frac{d^2 \kvec_{1 \perp}}{(2\pi)^2}
\, \frac{1}{k_{1 \perp}^2 k_{2 \perp}^2} \nonumber \label{shorthandB} \\
J(k_1^-) &=& \frac{1}{k_1^- + k_{1 \perp}^2/M - i\epsilon} +
\frac{1}{k_2^- + k_{2 \perp}^2/M - i\epsilon}
\eeqa
In order to isolate the poles at $k_i^+=0$ coming from the gluon
propagators we view the integrands in \eq{A3} as rational functions
of $k_2^+$, which we decompose in terms of simple elements. Also,
since $p_1^-$ is the largest scale we can approximate:
\beq
\frac{1}{k_2^+[-p_1^- k_2^+ + D(p_1)-D(p_2-k_1) +i\epsilon]} \simeq
\frac{1}{D(p_1)-D(p_2-k_1)} \left(\frac{1}{k_2^+} - \frac{1}{k_2^+
     -i\epsilon} \right)
\eeq
We also use
\beq
\frac{1}{k_1^+ k_2^+} = \frac{1}{k^+} \left( \frac{1}{k_1^+} +
\frac{1}{k_2^+} \right)
\eeq
to arrive at
\beqa \label{Bsimple}
B_a + B_b &=& -eg^4 QM p_2^- \int_{\perp} \int \frac{dk_1^-}{2i\pi}
J(k_1^-) \int \frac{dk_2^+}{\pi} \left[ \frac{1}{D(p_1)}
-\frac{1}{D(p_2-k_2)} \right] \nonumber \\
&\times& \left\{ \frac{1}{k_2^+ - [D(p_2)-D(p_2-k_2)]/p_2^-
     -i\epsilon} \right.
\nonumber \\
&\ & \left. + \frac{1}{k_1^+} \left(1-\frac{D(p_2-k_2)}{D(p_2)}\right)
- \frac{1}{k_2^+} \frac{D(p_2-k_2)}{D(p_2)} \right\}  \\
    B_c + B_d &=& -eg^4 QM p_2^- \int_{\perp} \int \frac{dk_1^-}{2i\pi}
    J(k_1^-) \int \frac{dk_2^+}{\pi} \left[ \frac{1}{D(p_2)}
-\frac{1}{D(p_2-k_1)} \right] \nonumber \\
&\times& \left\{ \frac{1}{k_2^+  -i\epsilon}
    - \frac{1}{k_2^+} \right\} \nonumber \\
    B_e &=& -eg^4 QM p_2^- \int_{\perp} \int \frac{dk_1^-}{2i\pi}
    J(k_1^-) \int \frac{dk_2^+}{\pi} \frac{- \kvec_{1 \perp} \cdot
\kvec_{2 \perp}}{D(p_1) D(p_2)}
    \left\{ \frac{1}{k_1^+} + \frac{1}{k_2^+} \right\}  \nonumber
\eeqa
Using the relation
\beq
\label{k1k2}
2 \kvec_{1 \perp} \cdot \kvec_{2 \perp} = D(p_1) +  D(p_2) -
D(p_2-k_1) - D(p_2-k_2) \nonumber
\eeq
one easily checks that the terms
$\propto 1/k_i^+$ in \eq{Bsimple} give the contribution
\beqa
\label{gaugepoles}
&\ & -eg^4 QM p_2^- \int_{\perp}
\int \frac{dk_1^-}{2i\pi} \, J(k_1^-) \int \frac{dk_2^+}{2\pi} \\
&\times& \left\{ \frac{1}{k_1^+} \left[ \frac{1}{D(p_2)}
-\frac{2}{D(p_2-k_2)} +\frac{1}{D(p_1)} +
\frac{D(p_2-k_1)-D(p_2-k_2)}{D(p_1)D(p_2)} \right]  \right. \nonumber \\
&-&\left. \frac{1}{k_2^+} \left[ \frac{1}{D(p_2)}
-\frac{2}{D(p_2-k_1)} +\frac{1}{D(p_1)} +
\frac{D(p_2-k_2)-D(p_2-k_1)}{D(p_1)D(p_2)} \right]  \right\} \nonumber
\eeqa
which vanishes by symmetry of $\int_{\perp}$ and $J(k_1^-)$ under
$(k_1^+,k_1^-,\kvec_{1 \perp}) \leftrightarrow (k_2^+,k_2^-,\kvec_{2 \perp})$.
As a consequence, the sum of all diagrams is independent of the way one
regularizes the {\it spurious} poles at $k_i^+=0$. Noting that
\beq
\label{intcalp}
\int \frac{dk_1^-}{2i\pi} J(k_1^-) = 1
\eeq
the prescription independent result for $B$ reads
\beq
\label{prescindB}
B = -i eg^4 QM p_2^- \int_{\perp}  \left[ \frac{1}{D(p_2)}
-\frac{2}{D(p_2-k_2)} +\frac{1}{D(p_1)}  \right]
\eeq
in agreement with the result \eq{Bexpr} in Feynman gauge (and large $M$).

As an individual diagram may contain pole terms $\sim 1/k_i^+$, its value can
depend on the prescription. As an illustration, we give the expressions of the
different diagrams using the three following prescriptions:
\beq
\label{prescriptions}
\frac{1}{k_i^+} \rightarrow \left[\frac{1}{k_i^+} \right]_{\eta_i} =
\left\{
\begin{array}{cc}
k_i^+\left[(k_i^+ -i\eta_i)(k_i^+ +i\eta_i)\right]^{-1}  & ({\rm PV}) \\
\left[k_i^+ -i\eta_i\right]^{-1} & ({\rm K}) \\
\left[k_i^+ -i\eta_i \epsilon(k_i^-)\right]^{-1} & ({\rm ML})
\end{array} \right.
\eeq
namely the principal-value, Kovchegov\footnote{Only the $d^{- \perp}$ component
of the gauge field propagator in Eq. (4) of \cite{kov} contributes in our
calculation.} \cite{kov} and Mandelstam-Leibbrandt \cite{gl} prescriptions.
The `sign function' is denoted $\epsilon(x)=\Theta(x)-\Theta(-x)$. With the PV
prescription we have
\beq
\int dk_2^+ \left[\frac{1}{k_2^+} \right]_{\eta_2} = 0
\eeq
and get
\beqa \label{BPV}
B_a + B_b &=& -i eg^4 QM p_2^- \int_{\perp}  \left[ \frac{1}{D(p_1)}
-\frac{1}{D(p_2-k_2)} \right] \nonumber \\
B_c + B_d &=& -i eg^4 QM p_2^- \int_{\perp}  \left[ \frac{1}{D(p_2)}
-\frac{1}{D(p_2-k_1)} \right] \nonumber \\
    B_e &=& 0
\eeqa
Using the K prescription we obtain
\beqa \label{imlres}
B_a + B_b &=& -2 i eg^4 QM p_2^- \int_{\perp}  \left[ \frac{1}{D(p_1)}
-\frac{1}{D(p_2-k_2)} \right] \left[1-\frac{D(p_2-k_2)}{D(p_2)} \right]
\nonumber \\
B_c + B_d &=& 0 \\
    B_e &=& -i eg^4 QM p_2^- \int_{\perp} \frac{-2 \kvec_{1 \perp}
\cdot \kvec_{2
\perp} }{D(p_1) D(p_2)}  \nonumber
\eeqa
The calculation with the ML prescription is a little more complicated. Defining
\beqa \label{I1I2def}
I_1 &=& \int dk_1^- J(k_1^-) \Theta (- k_2^-) \nonumber \\
I_2 &=& \int dk_1^- J(k_1^-) \Theta (k_1^-)
\eeqa
and using \eq{intcalp} we get after regularizing \eq{Bsimple}
\beqa \label{BML}
B_a + B_b &=& -\frac{eg^4 QM p_2^-}{\pi} \int_{\perp}  \left[ \frac{1}{D(p_1)}
-\frac{1}{D(p_2-k_2)} \right] \nonumber \\
&\times& \left[I_1 \frac{D(p_2-k_2)}{D(p_2)} +
I_2 \left( 1-\frac{D(p_2-k_2)}{D(p_2)}\right) \right] \nonumber \\
B_c + B_d &=& -\frac{eg^4 QM p_2^-}{\pi} \int_{\perp}  \left[ \frac{1}{D(p_2)}
-\frac{1}{D(p_2-k_1)} \right] I_1 \nonumber \\
    B_e &=&  -\frac{eg^4 QM p_2^-}{\pi} \int_{\perp} \frac{\kvec_{1 \perp} \cdot
\kvec_{2 \perp} }{D(p_1) D(p_2)} \, (I_1-I_2)
\eeqa
Calculating explicitly $I_1$ and $I_2$ gives
\beqa \label{I1I2}
I_1 &=& \log\left(\frac{k_{2 \perp}^2}{|k_{1
\perp}^2-k_{\perp}^2|}\right) + i \pi
[1+ \Theta(k_{\perp}^2-k_{1 \perp}^2)] \nonumber \\
I_2 &=& \log\left(\frac{|k_{\perp}^2-k_{2 \perp}^2|}{k_{1
       \perp}^2}\right) + i \pi \Theta(k_{2 \perp}^2-k_{\perp}^2)
\eeqa
We can then use the relation
\beq
I_1(k_1,k_2) + I_2(k_2,k_1) = 2i \pi  \nonumber
\eeq
to check that the sum of all diagrams evaluated with the ML prescription indeed
reproduces the result \eq{prescindB}.

Instead of using \eq{Bsimple}, one can also directly use \eq{A3}, after
regularizing the $k_i^+=0$ poles with a chosen prescription (for
instance one of
those given in \eq{prescriptions}), and perform the $k_2^+$ integral using
Cauchy's theorem. The calculation is more involved, but reproduces all results
presented above. See the comments at the end of Appendix B concerning this
procedure.

\vspace{.5cm}

\centerline{{\bf B. Two-loop calculation in $A^+=0$ gauge.}}

\begin{figure}[htb]
\begin{center}
\leavevmode
{\epsfxsize=13.5truecm \epsfbox{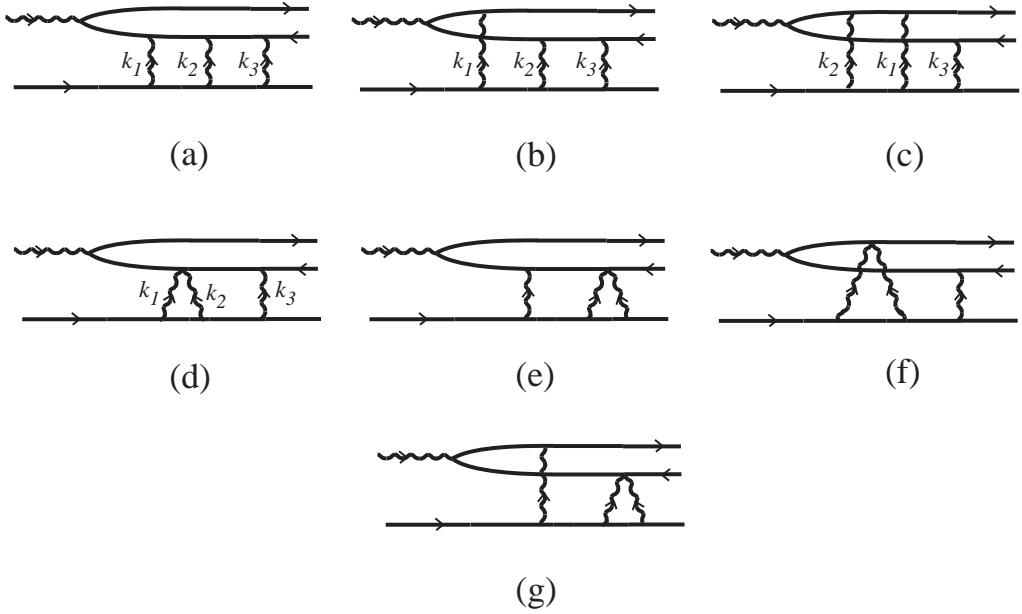}}
\end{center}
\caption[*]{Diagrams that can contribute to the two-loop amplitude $C$ in
$A^+=0$ gauge. All permutations of the attachments to the target line are
implied.}
\label{fig8dis}
\end{figure}
Here we evaluate the three-gluon exchange amplitude $C$ \eq{Cexpr} in
$A^+=0$ gauge and in the kinematic limit \eq{scales}. The leading
order diagrams
$C_a \ldots C_g$ are displayed in Fig. 8. For each diagram, the $6$
permutations
of the vertices on the target line are taken into account. Since two
permutations correspond to the same topology for diagrams $C_d \ldots C_g$,
there is a factor $1/2$ for those diagrams. We will use the following shorthand
notations:
\beqa
&\ &\int_{\perp} \equiv \int \frac{d^2 \kvec_{1 \perp}}{(2\pi)^2}\,
\frac{d^2 \kvec_{2 \perp}}{(2\pi)^2}
\, \frac{1}{k_{1 \perp}^2 k_{2 \perp}^2 k_{3 \perp}^2}\ ;\
\int_{+} \equiv \int \frac{d k_1^+}{2\pi} \frac{d k_2^+}{2\pi}\ ;\
\int_{-} \equiv \int \frac{d k_1^-}{2\pi i} \frac{d k_2^-}{2\pi i}
\nonumber \label{shorthandC} \\
&J& = \frac{1}{\left[k_1^- + k_{1 \perp}^2/M -
    i\epsilon\right] \left[k_1^- + k_2^- +
(\kvec_{1 \perp} + \kvec_{2 \perp})^2/M - i\epsilon\right] } +
{\rm perm}\,(k_1,k_2,k_3) \nonumber \\
\eeqa
(where $J$ contains $6$ terms arising from the $6$ permutations mentioned
above),
and
\beq
D_{ij} \equiv D(\pvec_{i \perp} - \kvec_{j \perp})\ ;\
D_{i} \equiv D(\pvec_{i \perp})\ ;\
{\rm for}\  i=1,2 \  {\rm and}\  j=1,2,3
\eeq
where $D$ is defined in \eq{Ddef}. Using the kinematic limit \eq{scales} and
approximations as in \eq{A1} and \eq{A2}, the scalar abelian Feynman
rules give:
\beqa
C_a &=& - 2eg^6 QM \int_{\perp} \int_{-} J \int_{+}
\frac{N_1 N_2 N_3}{k_1^+ k_2^+ k_3^+}
\frac{1}{-D_1 (p_2^- k_1^+ - D_{11} + i\epsilon)
(-p_2^- k_3^+ - N_3 + i\epsilon)} \nonumber \\
C_b &=& - 2eg^6 QM \int_{\perp} \int_{-} J \int_{+}
\frac{N_1 N_2 N_3}{k_1^+ k_2^+ k_3^+}  \nonumber \\
&\times& \frac{1}{(-p_1^- k_1^+ + N_1 + i\epsilon)
(p_2^- k_1^+ - D_{11} + i\epsilon)
(-p_2^- k_3^+ - N_3 + i\epsilon)} \nonumber \\
C_c &=& - 2eg^6 QM \int_{\perp} \int_{-} J \int_{+}
\frac{N_1 N_2 N_3}{k_1^+ k_2^+ k_3^+}  \nonumber \\
&\times& \frac{1}{(-p_1^- k_1^+ + N_1 + i\epsilon)
(- p_1^- (k_1^+ + k_2^+)  + N_1 + N_2 + i\epsilon)
(-p_2^- k_3^+ - N_3 + i\epsilon)} \nonumber \\
C_d &=& - 2eg^6 QM \int_{\perp} \int_{-} J \int_{+}
\frac{N_3 \, \kvec_{1 \perp} \cdot \kvec_{2 \perp}}{k_1^+ k_2^+ k_3^+}
\frac{1}{-D_1 (-p_2^- k_3^+ - N_3 + i\epsilon)} \nonumber \\
C_e &=& - 2eg^6 QM \int_{\perp} \int_{-} J \int_{+}
\frac{N_1 \, \kvec_{2 \perp} \cdot \kvec_{3 \perp}}{k_1^+ k_2^+ k_3^+}
\frac{1}{-D_1 (p_2^- k_1^+ - D_{11} + i\epsilon)} \nonumber \\
C_f &=& - 2eg^6 QM \int_{\perp} \int_{-} J \int_{+}
\frac{N_3 \, \kvec_{1 \perp} \cdot \kvec_{2 \perp}}{k_1^+ k_2^+ k_3^+}
\nonumber \\
&\times& \frac{1}{(- p_1^- (k_1^+ + k_2^+)  + N_1 + N_2 + i\epsilon)
(-p_2^- k_3^+ - N_3 + i\epsilon)} \nonumber \\
C_g &=& - 2eg^6 QM \int_{\perp} \int_{-} J \int_{+}
\frac{N_1 \, \kvec_{2 \perp} \cdot \kvec_{3 \perp}}{k_1^+ k_2^+ k_3^+}
\frac{1}{(-p_1^- k_1^+ + N_1 + i\epsilon)
(p_2^- k_1^+ - D_{11} + i\epsilon)} \nonumber \\
\label{Cdiag}
\eeqa
with
\beq
N_1=D_1 - D_{11}\ ;\  N_2=D_{11} - D_{23}\ ;\
N_3=D_{23} - D_2
\eeq

Similarly to what was done in Appendix A for the one-loop calculation,
one now considers all integrands in \eq{Cdiag} as rational functions
of $k_i^+$ ($i=1,2,3$), which we decompose in simple elements, making
first use of
\beq
\frac{1}{k_1^+ k_2^+ k_3^+} = \frac{1}{k^+} \left( \frac{1}{k_1^+ k_2^+} +
\frac{1}{k_1^+ k_3^+} + \frac{1}{k_2^+ k_3^+}\right)
\eeq
The limit $p_1^- \to \infty$ must be taken {\it after} the
decomposition in simple elements has been completed, otherwise some
pinch singularities can arise. As there are in the two-loop case two
independent `$+$' integration variables ($k_1^+ +  k_2^+ + k_3^+ =
k^+$), each integrand can be expressed as a sum of terms having one
of the following forms:
\beq
\label{forms}
\frac{1}{k_i^+ k_j^+}\ ;\ \frac{1}{k_i^+ (k_j^+ \pm i\epsilon)}\ ;\
\frac{1}{(k_i^+ \pm i\epsilon) (k_j^+ \pm i\epsilon)} \ \ \
(i \neq j)
\eeq
In \eq{forms} the poles at $k_i^+ =0$ come from the gluon propagators
in LC $A^+=0$ gauge, whereas those at $k_i^+ = \pm i\epsilon$
originate from the scalar quark propagators. The result of the full
decomposition is
\beqa
C_a &=& - 2eg^6 QMp_2^- \int_{\perp} \int_{-} J \int_{+}
\nonumber \\
&\times&
\left\{\left(\frac{1}{D_{23}}-\frac{1}{D_{11}}\right)
\left(1-\frac{D_{11}}{D_1}\right)
\left(\frac{D_{23}}{D_2}-1\right) \left[\frac{1}{k_1^+ + i\epsilon} -
\frac{1}{k_1^+}\right] \left[\frac{1}{k_2^+} +
\frac{1}{k_3^+ - i\epsilon} \right]  \right. \nonumber \\
&+& \left. \left(\frac{1}{D_{11}}-\frac{1}{D_1}\right)
\frac{D_{11}-D_{23}}{D_2}
\left[\frac{1}{k_1^+ + i\epsilon} - \frac{1}{k_1^+}\right]
\left[\frac{1}{k_3^+} - \frac{1}{k_3^+ - i\epsilon} \right]
\right. \nonumber \\
&+& \left. \frac{(D_1-D_{11})(D_{11}-D_{23})}{D_1 D_2 (D_2-D_{11})}
\left[\frac{1}{k_1^+ + i\epsilon} + \frac{1}{k_2^+}\right]
\left[\frac{1}{k_3^+} - \frac{1}{k_3^+ - i\epsilon} \right] \right\}
    \nonumber \\
C_b &=& - 2eg^6 QMp_2^- \int_{\perp} \int_{-} J \int_{+}
\nonumber \\
&\times&
\left\{\left(\frac{1}{D_2}-\frac{1}{D_{23}}\right)
\left(1-\frac{D_{23}}{D_{11}}\right)
   \left[\frac{1}{k_2^+} + \frac{1}{k_3^+ - i\epsilon} \right]
\left[\frac{1}{k_1^+} - \frac{1}{k_1^+ - i\epsilon} \right]
\right. \nonumber \\
&+& \left. \frac{1}{D_2}\left(1-\frac{D_{23}}{D_{11}}\right)
\left[\frac{1}{k_3^+} - \frac{1}{k_3^+ - i\epsilon} \right]
\left[\frac{1}{k_1^+} - \frac{1}{k_1^+ - i\epsilon} \right] \right\}
    \nonumber \\
C_c &=& - 2eg^6 QMp_2^- \int_{\perp} \int_{-} J \int_{+}
\nonumber \\
&\times&
\left(\frac{1}{D_{23}}-\frac{1}{D_2}\right)
\frac{D_{11}-D_{23}}{D_1-D_{23}}
\left[\frac{1}{k_2^+} + \frac{1}{k_3^+ + i\epsilon} \right]
\left[\frac{1}{k_1^+} - \frac{1}{k_1^+ - i\epsilon} \right]
    \nonumber \\
C_d &=& - 2eg^6 QMp_2^- \int_{\perp} \int_{-} J \int_{+}
\frac{\kvec_{1 \perp} \cdot \kvec_{2 \perp}}{D_1 D_2}
   \left\{ \left[\frac{1}{k_1^+} +  \frac{1}{k_2^+}\right]
\left[\frac{1}{k_3^+} - \frac{1}{k_3^+ - i\epsilon} \right]\right.
\nonumber \\
&+&\left.\left(1-\frac{D_2}{D_{23}}\right)
\left[\frac{1}{k_1^+} - \frac{1}{k_1^+ + i\epsilon} \right]
\left[\frac{1}{k_2^+} + \frac{1}{k_3^+ - i\epsilon} \right] \right\}
    \nonumber \\
C_e &=& - 2eg^6 QMp_2^- \int_{\perp} \int_{-} J \int_{+}
\frac{\kvec_{2 \perp} \cdot \kvec_{3 \perp}}{D_1 D_2}
\left\{ \left(1-\frac{D_1}{D_{11}}\right)
   \left[\frac{1}{k_2^+} +  \frac{1}{k_3^+}\right]
\left[\frac{1}{k_1^+ + i\epsilon} - \frac{1}{k_1^+} \right]\right.
\nonumber \\
&+& \left.\frac{D_{11}-D_1}{D_2-D_{11}}
\left[\frac{1}{k_2^+} - \frac{1}{k_2^+ - i\epsilon} \right]
\left[\frac{1}{k_3^+} + \frac{1}{k_1^+ + i\epsilon} \right]
\right\}\nonumber \\
C_f &=& - 2eg^6 QMp_2^- \int_{\perp} \int_{-} J \int_{+}
\frac{\kvec_{1 \perp} \cdot \kvec_{2 \perp}}{D_1- D_{23}}
\nonumber \\
&\times& \left(\frac{1}{D_2}-\frac{1}{D_{23}}\right)
\left[\frac{1}{k_1^+ - i\epsilon} - \frac{1}{k_1^+} \right]
\left[\frac{1}{k_2^+} + \frac{1}{k_3^+ + i\epsilon} \right]
\nonumber \\
C_g &=& - 2eg^6 QMp_2^- \int_{\perp} \int_{-} J \int_{+}
\frac{- \kvec_{2 \perp} \cdot \kvec_{3 \perp}}{D_2 D_{11}}
\left[\frac{1}{k_2^+} + \frac{1}{k_3^+} \right]
\left[\frac{1}{k_1^+} - \frac{1}{k_1^+ - i\epsilon} \right]
\label{Csimple}
\eeqa
Eq. \eq{Csimple} can be conveniently used to group together the poles
at $k_i^+=0$, which appear in the two first forms of \eq{forms}. For
each of these forms, a lengthy calculation shows that the $k_i^+=0$
poles add to a contribution which is identically zero, analogously to
\eq{gaugepoles} for the one-loop calculation. On the way we use the
identities
\beqa
\label{identity1}
-2 \kvec_{1 \perp} \cdot \kvec_{2 \perp} = D_{11} + D_{12} - D_{23} -
D_1  \nonumber \\
-2 \kvec_{2 \perp} \cdot \kvec_{3 \perp} = D_{13} + D_{12} - D_{21} - D_1
\eeqa
\beq
\label{identity2}
D_2 + D_{11} + D_{12} + D_{13} - D_{21} - D_{22} - D_{23} - D_1 =0
\eeq
and realize that in every factor
$[1/k_i^+ \pm 1/(k_j^+ \pm i\epsilon)]$ ($i \neq j$) of
\eq{Csimple}, $k_j^+$ can be replaced by $k_i^+$ (and not the
contrary\footnote{We do not allow the inverse change $1/k_i^+
    \rightarrow 1/k_j^+$ to keep the possibility to deal with a
    regularized form of $1/k_i^+$ depending on $k_i^-$, as is the case
    for the ML prescription, see \eq{prescriptions}.}) by a change of variable.
We also use the symmetry of $\int_{\perp}$ and $J$ under
$k_i \leftrightarrow k_j$ for $i \neq j$.

Thus we have explicitly checked the complete prescription independence
of our two-loop calculation. Only terms of the last form of \eq{forms}
remain in \eq{Csimple}. Using
\beq \label{doubleintJ}
\int_{-} J(k_1^-, k_2^-) = 1
\eeq
as well as \eq{identity1}, \eq{identity2} and symmetry arguments, one
shows that these terms add to
\beq
C = -\frac{1}{3} eg^6 M Q p_2^- \int_{\perp}
\left[ \frac{1}{D_2} -\frac{3}{D_{23}}
+ \frac{3}{D_{11}} - \frac{1}{D_1} \right]
\eeq
which exactly reproduces the result \eq{Cexpr} obtained in Feynman
gauge (for large $M$).

After having shown the complete prescription independence of our
calculation, we
conclude this Appendix with some important remarks. We stress that \eq{Cdiag}
and \eq{Csimple} are equivalent mathematical expressions for any of
the diagrams
$C_a \ldots C_g$.  To evaluate a given diagram, one needs to regularize the
$k_i^+=0$ poles, but this can be done starting either from \eq{Cdiag} or from
\eq{Csimple}, and the same results must follow. We have checked this for all
diagrams using the PV and K prescriptions. We thus see no problems in applying
the PV prescription to two-loop diagrams. Using the PV prescription on
\eq{Csimple} is straightforward, but applying it to \eq{Cdiag} requires some
comments. Regularizing
$1/(k_1^+ k_2^+ k_3^+)$ yields
\beq
\frac{1}{k_1^+ k_2^+ k_3^+} \rightarrow
\prod_{i=1}^{3}{\rm PV}\left(\frac{1}{k_i^+} \right)
   = \lim_{\eta_3 \to 0} \lim_{\eta_2 \to 0} \lim_{\eta_1 \to 0}
\left[\frac{1}{k_1^+} \right]_{\eta_1} \left[\frac{1}{k_2^+} \right]_{\eta_2}
\left[\frac{1}{k_3^+} \right]_{\eta_3}
\eeq
where $[1/k_i^+]_{\eta_i}$ is given in \eq{prescriptions}. Thus the poles at
$k_i^+=0$ must be regularized with {\it distinct} small finite parameters
$\eta_i$. Then the $k_i^+$ integrals are performed using Cauchy's theorem, and
only in the end the limits $\eta_1 \to 0$, $\eta_2 \to 0$, $\eta_3 \to 0$ are
taken  separately (in arbitrary order). We found this procedure to be
well-defined and to give results consistent with those directly obtained from
\eq{Csimple}.

Finally, as in the one-loop case, it is remarkable that the K
prescription makes all  two-loop diagrams where the fast quark
rescatters vanish, \ie, only $C_a$, $C_d$ and $C_e$ contribute to
the amplitude $C$.

\vspace{.5cm}

\centerline{{\bf C. Absence of struck quark rescattering in $A^+=0$ gauge.}}

In this Appendix we show that in $A^+=0$ gauge, independently of
the prescription used to regularize the {\it spurious} $k_i^+=0$
poles, rescatterings of the struck quark $p_1$ cancel in the cross
section, \ie \  after summing over cuts
in the forward Compton amplitude.
This is done by proving that the
full contribution to the cross section (use Eqs. \eq{Atildeexpr},
\eq{Btildeexpr}, \eq{Ctildeexpr})
\beq
\label{fullcrosssection}
\int d^2\rvec_\perp\, d^2\Rvec_\perp\, \left[ |{\tilde B}|^2 + 2
{\tilde A} {\tilde C} \right] = - \frac{1}{3} (e g^4 Q M p_2^-)^2
\int d^2\rvec_\perp\, d^2\Rvec_\perp\,
V(m_\pl r_\perp)^2 W(\rvec_\perp, \Rvec_\perp)^4
\eeq
is given by attachments to $p_2$ only.

We need to know the
partial amplitudes $A_2$, $B_2$, $C_2$ contributing to $A$,
$B$, $C$ where only attachments to $p_2$ are kept. For the Born
amplitude $A$, only the diagram of Fig. 4a contributes in $A^+=0$
gauge. Thus the partial amplitude $A_2$ from attachments to $p_2$
is actually the full amplitude $A$ given in \eq{Aexpr},
\beq
A_2 = A = \frac{2eg^2 M Q p_2^-}{k_\perp^2}
\left[\frac{1}{D_2} - \frac{1}{D_1} \right]  \label{calAexpr}
\eeq
The partial one-loop amplitude $B_2$ is given by the sum of the
diagrams $B_a$, $B_b$ and $B_e$ of Fig. 7. This sum is prescription
dependent.
We use the notation (see Eq. \eq{prescriptions})
\beq
I_{\eta} = \int \frac{dk_2^+}{i \pi} \left[\frac{1}{k_2^+} \right]_{\eta}
\eeq
giving $I_{PV}=0$ and $I_{K}=1$.
After regularizing \eq{Bsimple} and using \eq{intcalp} and symmetry
arguments
we find\footnote{In order to use \eq{intcalp} in \eq{Bsimple}, we need
   to consider a regularized form of $1/k_i^+$ independent of $k_i^-$
   (\ie \  we exclude for simplicity the ML prescription in this
   Appendix). Eq. \eq{calBexpr} is valid for any such prescription.},
\beq
B_2 = B_a + B_b + B_e = -i \, e g^4 Q M p_2^-
\int_{\perp} \left[\left( \frac{1}{D_1} - \frac{1}{D_{22}}\right) +
I_{\eta} \, \left( \frac{1}{D_2} - \frac{1}{D_{22}} \right)   \right]
\label{calBexpr}
\eeq
The partial two-loop amplitude $C_2$ is given by the diagrams
$C_a$, $C_d$ and $C_e$ of Fig. 8. Regularizing \eq{Csimple} and using
\eq{doubleintJ} we get
\beqa
C_a &=& - \frac{1}{2} eg^6 QMp_2^- \int_{\perp} \,
\left\{\left(\frac{1}{D_{23}}-\frac{1}{D_{11}}\right)
\left(1-\frac{D_{11}}{D_1}\right)
\left(\frac{D_{23}}{D_2}-1\right) (I_{\eta} + 1)^2 \right. \nonumber \\
&+& \left. \left(\frac{1}{D_{11}}-\frac{1}{D_1}\right)
\frac{D_{11}-D_{23}}{D_2} (I_{\eta}^2 - 1) -
\frac{(D_1-D_{11})(D_{11}-D_{23})}{D_1 D_2 (D_2-D_{11})}
(I_{\eta} - 1)^2  \right\}
   \nonumber \\
C_d &=& \frac{1}{2} eg^6 QMp_2^- \int_{\perp}
\frac{\kvec_{1 \perp} \cdot \kvec_{2 \perp}}{D_1 D_2}
  \left\{2 I_{\eta} (I_{\eta} - 1)
+ \left(1-\frac{D_2}{D_{23}}\right) (I_{\eta} + 1)^2 \right\}
   \nonumber \\
C_e &=& \frac{1}{2} eg^6 QMp_2^- \int_{\perp}
\frac{\kvec_{2 \perp} \cdot \kvec_{3 \perp}}{D_1 D_2}
\left\{ \left(1-\frac{D_1}{D_{11}}\right) 2 I_{\eta} (- I_{\eta} - 1)
+ \frac{D_{11}-D_1}{D_2-D_{11}} (I_{\eta} - 1)^2
\right\}  \nonumber \\
\label{Cade}
\eeqa
Using \eq{identity1}, \eq{identity2} and symmetry arguments we get
after some algebra
\beqa
C_2 &=& C_a + C_d + C_e = - \frac{1}{2} eg^6 QMp_2^-
\int_{\perp} \nonumber \\
&& \left\{
   \left[-\frac{2}{3D_1}+\frac{1}{6D_2}-\frac{1}{2D_{23}}+\frac{1}{D_{11}}
\right] +
I_{\eta} \left[-\frac{1}{D_{23}}+\frac{1}{D_{11}} \right]
+ I_{\eta}^2 \left[\frac{1}{2D_{2}}-\frac{1}{2D_{23}} \right] \right\}
\nonumber \\
\label{calCexpr}
\eeqa
It is an easy exercise to express the partial amplitudes
$A_2$, $B_2$, $C_2$ in transverse coordinate space,
as was done for the full amplitudes $A$, $B$, $C$ in section 5
(see Eq. \eq{Atilde}).
Since the partial amplitudes are not infrared finite\footnote{Note
   however that with the K prescription ($I_{\eta}=1$)
   the partial
   amplitudes \eq{tildecalABC} equal the full ones, as already
   mentioned at
   the end of Appendix B. Thus the partial amplitudes are
   finite when $\lambda \to 0$ with this particular prescription.},
we introduce a small photon mass $\lambda$ in the exchanged photon
propagators, \ie \  $1/k_{i\perp}^2 \rightarrow 1/(k_{i\perp}^2 + \lambda^2)$
in the definition of $\int_{\perp}$ \eq{shorthandB} or \eq{shorthandC}.
Then
\beqa
{\tilde A_2} &=& \ 2\, eg^2 QMp_2^- V W \nonumber \\
{\tilde B_2} &=& -i\, eg^4 QMp_2^- V W \,
\frac{I_{\eta} K_0(\lambda R_{\perp}) -
K_0(\lambda |\Rvec_\perp+\rvec_\perp|)}{2\pi}
\nonumber \\
{\tilde C_2} &=& - \frac{1}{4}\, eg^6 QMp_2^- V W
\left[ \frac{W^2}{3} +  \left( \frac{I_{\eta} K_0(\lambda R_{\perp}) -
K_0(\lambda |\Rvec_\perp+\rvec_\perp|)}{2\pi}\right)^2 \right]
\label{tildecalABC}
\eeqa
where $V$ and $W$ stand for $V(m_\pl r_\perp)$ and
$W(\rvec_\perp, \Rvec_\perp)$.
The contribution from attachments to $p_2$ to the cross section reads
\beq
\int d^2\rvec_\perp\, d^2\Rvec_\perp\, \left[ |{\tilde B_2}|^2 + 2
{\tilde A_2} {\tilde C_2} \right] = - \frac{1}{3} (e g^4 Q M p_2^-)^2
\int d^2\rvec_\perp\, d^2\Rvec_\perp\, V^2 W ^4
\label{crosssectionfromp2}
\eeq
This is infrared finite, prescription independent, and identical to
the full result \eq{fullcrosssection}. Hence rescatterings of the
struck quark $p_1$ cancel in the cross section in $A^+=0$ gauge.

  From \eq{fullcrosssection} and \eq{crosssectionfromp2} we see
that in $A^+=0$ gauge and in coordinate space, the
contribution from attachments to $p_2$ equals the full contribution
$|B|^2 + 2 AC$ even at the integrand level, \ie \  before integrating
over $\rvec_\perp$ and $\Rvec_\perp$. We thus have, in coordinate space,
\beq
|{\tilde {\cal M}}|^2 = |{\tilde {\cal M}_2}|^2
\eeq
where ${\tilde {\cal M}} = \tilde A + \tilde B + \tilde C + \ldots$
is given in (39) and
${\tilde {\cal M}_2} = \tilde A_2 + \tilde B_2 + \tilde C_2 + \ldots$
corresponds to the partial amplitude where only attachments to $p_2$
are kept. Thus in $A^+=0$ gauge
\beq
\label{phasediff1}
{\tilde {\cal M}}(\rvec_\perp,\Rvec_\perp) = e^{i \phi}
{\tilde {\cal M}_2}(\rvec_\perp,\Rvec_\perp)
\eeq
The full amplitude ${\tilde {\cal M}}$ is obtained from
${\tilde {\cal M}_2}$ by inserting any number of rescatterings of
the quark $p_1$.
  Eq. \eq{phasediff1} reads
\beq
\label{expansion}
\tilde A + \tilde B + \tilde C + \ldots = e^{i \phi} \,
(\tilde A_2 + \tilde B_2 + \tilde C_2 + \ldots)
\eeq
By expanding the l.h.s. and r.h.s. of \eq{expansion} up to order
$g^6$, one realizes that $\phi$
must be at least of order $g^2$,
\beq
\phi = \phi_1 \, g^2 + \phi_2 \, g^4 + \ldots
\eeq
Identification of the terms of order $g^4$ and $g^6$ in the two sides
of \eq{expansion} leads to
\beqa
\phi_1 &=& \frac{I_{\eta}-1}{4 \pi} \, K_0(\lambda R_{\perp}) \nonumber \\
\phi_2 &=& 0
\eeqa
or
\beq
\phi = g^2 \, \frac{I_{\eta}-1}{4 \pi} \, K_0(\lambda R_{\perp}) +
{\cal{O}}(g^6)
\eeq
Although not proven here, the \order{g^6} terms in $\phi$
are expected to
vanish because adding one rescattering of $p_1$ can only bring a
power $g^2$ (see also the following discussion). Thus we get
\beq
\label{phasediff2}
{\tilde {\cal M}}(\rvec_\perp,\Rvec_\perp) =
{\rm exp}{\left[i \frac{g^2}{4 \pi} (I_{\eta}-1) K_0(\lambda R_{\perp})
\right]} \, {\tilde {\cal M}_2}(\rvec_\perp,\Rvec_\perp)
\eeq
As expected, since ${\tilde {\cal M}}$ is infrared safe and
prescription independent, all the dependence on $\lambda$ and
$I_{\eta}$ of ${\tilde {\cal M}_2}$ is contained in the phase. Note
also that with the K prescription, $I_{\eta}=1$ and
${\tilde {\cal M}} = {\tilde {\cal M}_2}$.

Eq. \eq{phasediff2} can also be understood as follows. In momentum
space, if we call $\mu$ the Lorentz index associated to the coupling
of $p_1$, we know that the amplitude is dominated by the
$n^{\nu} k_i^{\mu} / k_i^{+}$ term of the
exchanged gluon propagator in $A^+=0$ gauge,
with $\mu = \perp$ and $\nu = -$. Together with the
scalar quark propagator
\beq
\Delta_i^{-1} \propto (p_1-k_i)^2 -m^2 + \ieps
\simeq - p_1^- k_i^{+} + a_i + \ieps
\eeq
where $a_i = D(p_1) - D(p_1-k_i)$, the factor $1/k_i^{+}$ yields
\beq
\label{factors}
\int \frac{dk_i^{+}}{2\pi i} \frac{1}{(- p_1^- k_i^{+} + a_i + \ieps
   )k_i^{+}} \rightarrow \frac{1}{a_i} \int \frac{dk_i^{+}}{2\pi i}
\left( \left[\frac{1}{k_i^+} \right]_{\eta}  + \frac{1}{- k_i^{+} +
     \ieps}  \right) = \frac{1}{a_i} \frac{I_{\eta}-1}{2}
\eeq
where some simplification similar to (58) was made. The scalar
coupling brings a factor
$g^2 \kvec_{i \perp} \cdot (2\pvec_1 - \kvec_i)_{\perp} = g^2 a_i$
which compensates the prefactor in the r.h.s of \eq{factors}.
We are left with the $1/(k_{i\perp}^2 + \lambda^2)$
factor from the gluon propagator, which after Fourier transform
gives $K_0(\lambda R_{\perp})/(2\pi)$.
This builds the complete phase in \eq{phasediff2}.

\end{document}